\documentclass[12pt,a4paper]{article}

\usepackage{color}
\usepackage{svg} 

\usepackage{caption}
\usepackage{subcaption}
\usepackage[a4paper, total={6.25in, 8in}]{geometry}

\usepackage{graphicx}
\usepackage{newtxtext}
\usepackage{newtxmath}
\usepackage{natbib}
\usepackage{hyperref}
 \hypersetup{
   colorlinks   = true, 
   urlcolor     = blue,
   linkcolor    = blue, 
   citecolor   = blue 
 }
\usepackage{amsmath}
\usepackage{ulem}
\usepackage{authblk}

\newcommand{\bbeta}{\hat{\boldsymbol y}}
\newcommand{\Dfour}{\mathbf{D}^4} 

\title{Propulsive performance of oscillating plates with time-periodic flexibility}

\author[1]{David Yudin\thanks{Email address for correspondence: davidy@udel.edu
}}
\author[2]{Daniel Floryan}
\author[1]{Tyler Van Buren}
\affil[1]{Mechanical Engineering, University of Delaware, Newark, Delaware 19716, USA}
\affil[2]{Mechanical Engineering, University of Houston, Houston, Texas 77204, USA}

\begin{document}

\maketitle
\begin{abstract}
We use small-amplitude inviscid theory to study the swimming performance of a flexible flapping plate with time-varying flexibility. The stiffness of the plate oscillates at twice the frequency of the kinematics in order to maintain a symmetric motion. Plates with constant and time-periodic stiffness are compared over a range of mean plate stiffness, oscillating stiffness amplitude, and oscillating stiffness phase for isolated heaving, isolated pitching, and combined leading edge kinematics. We find that there is a profound impact of oscillating stiffness on the thrust, with a lesser impact on propulsive efficiency. Thrust improvements of up to 35\% relative to a constant-stiffness plate are observed. For large enough frequencies and amplitudes of the stiffness oscillation, instabilities emerge. The unstable regions may confer enhanced propulsive performance; this hypothesis must be verified via experiments or nonlinear simulations. 

\end{abstract}

\section{Introduction}
For decades research has focused on studying the fluid dynamics of biological swimmers, both to better understand the underlying biology of aquatic animals, and to provide inspiration for developing innovative hydrodynamic propulsion technology \citep{smits_2019}. A salient feature of natural swimmers is the action of muscles, which are often distributed throughout an animal's propulsor \citep{flammang2008speed, adams2019odontocete}. Through observations and measurements of animals, we know that swimmers can control their fin curvature, displacement, and area as well as their stiffness \citep{doi:10.1146/annurev.fluid.38.050304.092201}. It stands to reason that swimmers may be able to utilize their muscles on the time scale of the oscillation of their propulsors to tune performance, e.g., by dynamically changing the stiffness of their propulsors; however, there is no definitive observation or consensus in the biological community that swimmers take advantage of their muscles in this way \citep{FishLauderPrivateComm}. In this work we will show that, from a purely hydrodynamic perspective, time-varying stiffness leads to propulsive benefits over constant-stiffness propulsors. 

The fluid dynamics of biological swimmers is rooted in the theory of rigid-wing flutter. 
\cite{theodorsen1935general} was first to theoretically model the forces produced by oscillating foils in a fluid, which was later extended by \cite{Garrick1936PropulsionOA} to analytically predict thrust and power for a rigid oscillating foil. Although these works focused on wing flutter, the connection to swimmers was clear. Later, \cite{chopra_kambe_1977} incorporated the effects of three-dimensionality via lifting line theory. \cite{anderson_streitlien_barrett_triantafyllou_1998} used particle image velocimetry to calculate thrust and power of harmonically oscillating foils and compared the results to inviscid theory predictions. More recently, \cite{floryan2017scaling} derived and experimentally validated propulsive scaling laws for rigid two-dimensional foils in pure heaving and pitching motions. This was extended to combined pitch-and-heave motions \citep{VanBurenAIAA} and non-sinusoidal motions \citep{VanBurenPRF, floryan2017forces}.

The passive flexibility (or elasticity) of an oscillating propulsor plays a key role in its propulsive performance. In a particularly influential work, \cite{wu_1961} was one of the first to analytically consider passive flexibility, albeit through prescribed kinematics.  \cite{katz_weihs_1978,katz_weihs_1979} calculated the coupled fluid-structure interactions for a two-dimensional flexible foil. Since then, many analytical, experimental, and computational studies have shown the influence of propulsor flexibility on characteristics like thrust and swimming efficiency, generally finding that flexibility can dramatically increase thrust and/or efficiency compared to a rigid propulsor \citep{alben_2008, doi:10.1063/1.3177356, dewey_boschitsch_moored_stone_smits_2013, quinn_lauder_smits_2014, doi:10.1063/1.4939499,moore_2014,Moore2015}.
In a particularly relevant work, \cite{floryan_rowley_2018} used small-amplitude theory to explore the role of resonance in constant-stiffness propulsors, finding that the benefits of resonance manifest in the thrust and power, but not necessarily in the propulsive efficiency. This analysis was extended in \cite{goza2020connections} to consider the role of nonlinearity in the fluid-structure system. 

Few works have explored beyond simple uniform and passive flexibility. \cite{floryan_rowley_2020} considered the role of stiffness distribution, finding that concentrating stiffness toward the leading edge produced higher thrust, but lower efficiency compared to foils with stiffness concentrated away from the leading edge. \cite{Quinn_2021} studied \emph{tuneable} stiffness---that is, quasi-steady changes in stiffness---where it was shown that stiffness could be tuned to maximize desired performance parameters. To our knowledge, the only work that considers time-varying stiffness at the same time scale as the kinematics is \cite{doi:10.1063/5.0027927} and \cite{hu2021effects}. The former used a finite-volume Navier-Stokes solver to study the effects of changing the flexibility of a nonlinear beam in the context of micro-air-vehicles, and the latter approximated a flexible plate by a rigid propulsor with a torsional spring at its leading edge heaving actively and pitching passively, respectively. By changing the stiffness of the torsional spring in time, \cite{hu2021effects} found that both the thrust and propulsive efficiency could be enhanced in swimming. \cite{doi:10.1063/5.0027927} finds there can be an increase in thrust of up to $52\%$, with little impact on efficiency.

In this work, we study the effects of time-periodic stiffness on the propulsive performance of a flexible plate. We solve a potential flow model that strongly couples the fluid and structural equations of motion. We use a pseudospectral method developed in \cite{MOORE2017792} to solve the equations of motion, introducing time-varying flexibility through a Fourier series expansion in time. This fast solver allows us to explore the propulsive performance over a large range of parameters. We then use Floquet theory to assess the stability of the system. We will show that the oscillating stiffness can strongly influence thrust and and weakly influence the efficiency of the oscillating propulsor, indicating that animals or human-made swimmers would benefit from this capability. 

\section{Problem statement and solution methods}

\begin{figure}
    \centering
    \includegraphics[width = 1\textwidth]{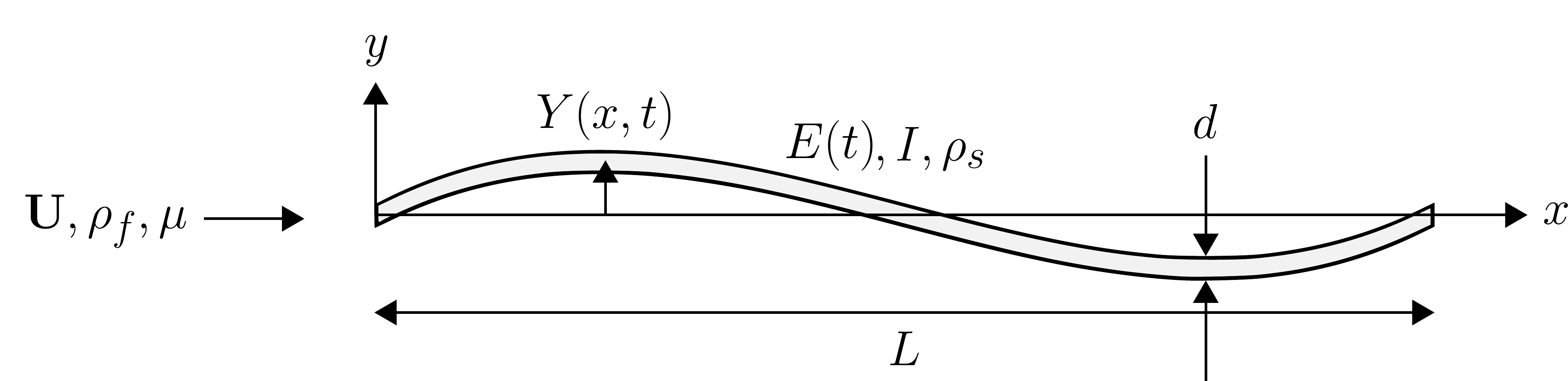}
    \caption{A two-dimensional flat plate with time-varying Young's modulus moving through a fluid.}
    \label{fig:schematic}
\end{figure}

Consider a two-dimensional, inextensible plate in a fluid flow, as sketched in Figure~\ref{fig:schematic}. The plate has thickness $d$ and length $L$, and we assume that it is thin ($d \ll L$) and that its maximum deflection is small, with its slope {$|Y_x| \ll 1$}. 
The deflection of the plate is then governed by the Euler-Bernoulli beam equation,

\begin{equation}
    \rho_s d w Y_{tt} + E(t)I Y_{xxxx} = w \Delta p,
    \label{Dimensional equation of motion}
\end{equation}
where $Y$ is the transverse displacement of the plate, $\rho_s$ is its density, $E(t)$ is its time-varying Young's modulus, $I = w d^3/12$ is its second moment of area, and $w$ is its width. The plate is immersed in a fluid which imparts a hydrodynamic load onto it, given by the pressure difference across the plate, $\Delta p$. Subscript $t$ denotes differentiation with respect to time, and subscript $x$ denotes differentiation with respect to the streamwise coordinate. 

The fluid is inviscid and incompressible, with density $\rho_f$. Far from the plate, the fluid moves with a freestream velocity $\boldsymbol U = U \boldsymbol i $, where $\boldsymbol i$ is the unit vector in the $x$ direction. Conservation of mass and momentum for the fluid lead to
\begin{subequations}
    \begin{gather}
        \boldsymbol \nabla \cdot \boldsymbol u = 0, \\
        \rho_f ( \boldsymbol u_t +  U \boldsymbol u _x) = - \boldsymbol \nabla p, \label{eq:mom}
    \end{gather}
\end{subequations}
where $\boldsymbol u = u \boldsymbol i + v \boldsymbol j$ is the perturbation velocity induced by the motion of the plate, and $\boldsymbol j$ is the unit vector in the transverse direction. In obtaining~\eqref{eq:mom}, we have assumed that $|\boldsymbol u| \ll U$, which is consistent with our assumption of small-amplitude deflections of the plate.

We nondimensionalize the equations of motion using the half-length of the plate $L/2$ as the length scale, the freestream velocity $U$ as the velocity scale, and the convective time $L/(2U)$ as the time scale. The nondimensional equation for the plate is 
\begin{equation}
        2RY_{tt}+ \frac{2}{3} S(t) Y_{xxxx} = \Delta p,
    \label{Non-dimensional equation of motion}
\end{equation}
and the nondimensional equations for the fluid are
\begin{subequations}
    \begin{gather}
        \boldsymbol \nabla \cdot \boldsymbol u = 0, \\
        \boldsymbol u_t +  \boldsymbol u _x = \boldsymbol \nabla \phi, \label{eq:ndmom}
    \end{gather}
\end{subequations}
where 
\begin{gather}
     R = \frac{\rho_s d}{\rho_f L},\;\;\; S(t) = \frac{E(t) d^3}{\rho_f U^2 L^3},\;\;\; \phi = p_\infty - p. 
     \label{Non-dimensional paramters}
\end{gather}
The function $\phi$ is Prandtl's acceleration potential \citep{wu_1961}. Now $x$, $t$, $Y$, $\boldsymbol u$, and $p$ are nondimensional, with $x = -1$ corresponding to the leading edge of the plate, and $x = 1$ corresponding to the trailing edge. The mass ratio $R$ is the ratio of a characteristic mass of the plate to a characteristic mass of fluid, and the stiffness ratio $S$ is the ratio of a characteristic bending force to a characteristic fluid force. The stiffness ratio, its inverse, and variations of it are sometimes called the Cauchy number \citep{ de2008effects} or the elastohydrodynamical number \citep{schouveiler2006rolling}. 

To close the system of equations, we specify the boundary conditions. The fluid satisfies the no-penetration boundary condition and the Kutta condition, 
\begin{subequations}
    \begin{gather}
    v|_{x \in [-1,1],y=0} = Y_{t} + Y_x, \label{eq:npbc}\\
    |v||_{(x,y) = (1,0)}  < \infty.
    \end{gather}
\end{subequations}
We specify heaving and pitching motions $h$ and $\theta$, respectively, at the leading edge of the plate, while the trailing edge is free (zero force and torque), resulting in the boundary conditions
\begin{equation}
        \label{eq:bcog}
        Y(-1,t) = h(t) , \;\;\; Y_{x}(-1,t) = \theta(t), \;\;\; Y_{xx}(1,t) = 0 ,\;\;\; Y_{xxx}(1,t) = 0.
\end{equation}

Since we are interested in locomotion, we consider periodic actuation of the leading edge; that is, $h(t)$ and $\theta(t)$ are (zero-mean) periodic functions of $t$ with a nondimensional angular frequency $\sigma = \pi f L / U$, where $f$ is the dimensional ordinary frequency. Similarly, we consider a time-periodic stiffness. With our focus on forward propulsion, the upstroke and downstroke must be mirror images of each other in order for the mean side force to be zero. It can be shown that this requires that the stiffness vary at twice the frequency of the deflection of the plate. Intuitively, the stiffness must be the same during the upstroke and downstroke, which can only be true if it varies at twice the frequency of the plate's deflection. 

To solve the system of equations for the kinematics of the plate, we use the method described by \citet{MOORE2017792}, adapting it to account for time-periodic stiffness; we describe the method in Appendix~\ref{ss: Solution method}. The method assumes that the kinematics are time-periodic, with any transients decaying to zero. We will test this assumption by performing a Floquet analysis. More importantly, the Floquet analysis will provide physical insight into the problem. The Floquet analysis is adapted from the eigenvalue problem described by \citet{floryan_rowley_2018, floryan_rowley_2020} and uses a key idea from \citet{kumar_tuckerman_1994}; the details are in Appendix~\ref{s:floquet theory}. 

The motion of the plate induces a pressure difference across it. The projection of the pressure difference onto the horizontal direction contributes---together with a suction force at the leading edge---to a propulsive thrust force. Therefore, the energy put into the system by actuating the leading edge is converted into propulsive thrust, given by
\begin{equation}
    C_T = \int_{-1}^{1}\Delta p Y_x dx + C_{TS},
    \label{CT eq}
\end{equation}
where $C_{TS}$ is the leading edge suction, a formula for which is given by \citet{MOORE2017792}. The power input is \begin{equation}
    C_P = -\int_{-1}^{1} \Delta p Y_t dx.
    \label{CP eq}
\end{equation}
Finally, the Froude efficiency is defined as the ratio of the time-averaged thrust output to the time-averaged power input (i.e., how much of the power input is converted to propulsive thrust),
\begin{equation}
    \eta = \frac{\overline{C_T}}{\overline{C_P}},
\end{equation}
where the overbar denotes averaging in time. 


In this work, we restrict ourselves to leading edge actuation and stiffnesses that are sinusoidal in time,
\begin{subequations}
\begin{align}
    h(t) &= \frac{1}{2}\left( h_0 e^{j \sigma t} + h_0^*e^{-j \sigma t}\right),
    \label{heave BC} \\
    \theta(t) &=  \frac{1}{2}\left(\theta_0 e^{j \sigma t} + \theta_0^* e^{-j \sigma t}\right),
    \label{pitch BC} \\
    S(t) &=  \overline S + \frac{\overline S}{2}\left(S_0 e^{2j \sigma t} + S_0^*e^{-2j \sigma t}\right) \label{eq:stif},
\end{align}
\end{subequations}
where $h_0, \theta_0, S_0 \in \mathbb{C}$, $j = \sqrt{-1}$, and a superscript $*$ denotes complex conjugation. The formulation in the appendices, however, is valid for generic smooth periodic functions of time. We will make frequent use of the parameter $\phi_S = \text{arg}(S_0)$, the phase of the stiffness oscillation. Note that $|S_0|$ gives the amplitude of the stiffness oscillation as a fraction of the mean stiffness; for example, $S_0 = 0.5$ means that the stiffness oscillates with an amplitude that is 50\% of the mean stiffness. For a physically meaningful (i.e., positive) stiffness, we require $|S_0| < 1$. 

Throughout, we fix the mass ratio to a low value of $R = 0.01$, appropriate for thin, neutrally buoyant biological swimmers. To build intuition for the effects of time-varying stiffness, we will extensively study the case with $\overline S = 20$; unless otherwise noted, this is the value we use for the mean stiffness. We will consider cases where the plate is in pure heave, in pure pitch, and in combined heave-and-pitch. Throughout, we set $h_0 = 1$ and $\theta_0 = 0.5$, though we will add a phase offset between heave and pitch for combined motions. 

\section{Results and discussion}

Heaving and pitching motions have fundamentally different thrust mechanisms \citep{floryan2017scaling, van2020bioinspired}, with thrust in heave coming from lift-based circulatory forces, and in pitch coming from added-mass acceleration forces. To start, we consider a periodically heaving plate moving through a fluid. After completing our heave-only analysis, we will analyze pitch-only and heave-and-pitch motions. 

\subsection{Heave-only motions}

\begin{figure}
    \centering
          \includegraphics[width = 1\textwidth]{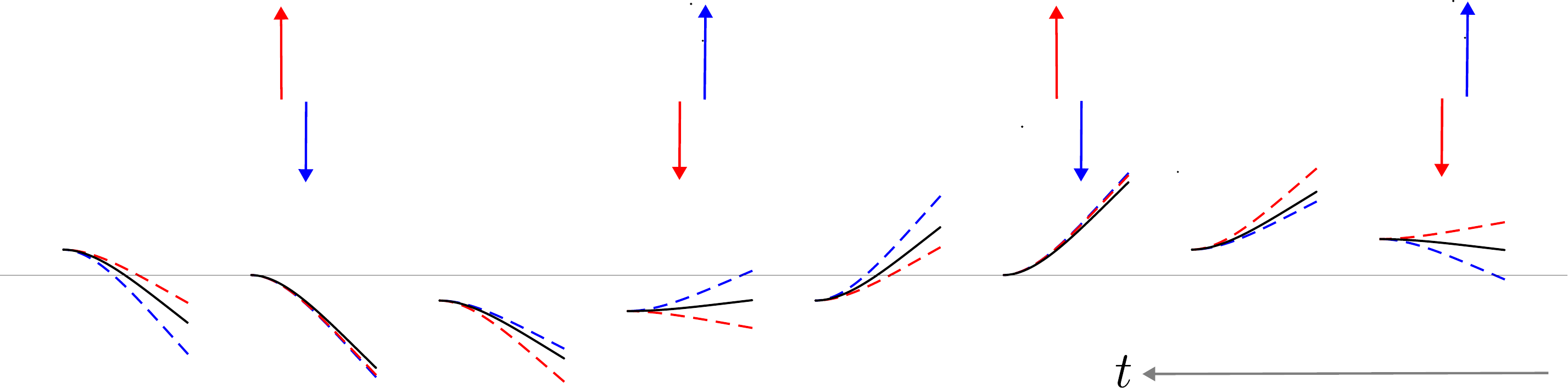}
   \caption{Kinematics of a heaving plate actuated at its first resonant frequency of $\sigma = 3.1$. Shown are a plate with constant stiffness (solid black), time-periodic stiffness in phase with the motion ($\phi_S = 0$; dashed blue), and time-periodic stiffness out of phase with the motion ($\phi_S = \pi$; dashed red). The colored arrows represent the difference between the instantaneous stiffness and the mean stiffness. The absence of arrows indicates the three plates have the same stiffness in the below snapshot. The parameters used are $\overline S = 20$, $h_0 = 1$, $R = 0.01$, and $|S_0| = 0.5$.}
    \label{fig:baselineHeaveKinematics}
\end{figure}

First, we familiarize ourselves with how time-varying flexibility impacts the plate's kinematics. Figure~\ref{fig:baselineHeaveKinematics} shows the kinematics of three plates during one cycle of motion. The plates are all actuated at the same frequency---the first resonant frequency of the plate with constant stiffness, $\sigma = 3.1$. The reference case with constant stiffness is compared against a plate with time-varying flexibility that is in phase with the motion ($\phi_S = 0$), meaning it is stiff at the turn-around and flexible at mid-stroke, and a plate with time-varying flexibility that is out of phase with the motion ($\phi_S = \pi$), meaning it is flexible at the turn-around and stiff at the mid-stroke. For a plate with constant stiffness, the peak deflection occurs around the mid-stroke, where the lateral velocity is highest. For the $\phi_S=0$ case, the pressure on the plate is reduced throughout the mid-stroke with more deflection, and the increase in stiffness towards the turnaround causes it to catch back up to the reference case. Conversely, the $\phi_S=\pi$ case has its stiffness reduced at the turnaround, ultimately achieving the largest trailing edge deflection of all heave cases, and then recovering throughout the mid-stroke with increased stiffness. The stiffness oscillation essentially adds a phase lag relative to the motion of the constant-stiffness plate. When $\phi_S = \pi$, the motion leads that of the constant-stiffness plate, whereas when $\phi_S = 0$, the motion lags that of the constant-stiffness plate. 

\begin{figure}
    \centering
    \begin{subfigure}{0.5\textwidth}
  \centering
  \includegraphics[width=1\linewidth]{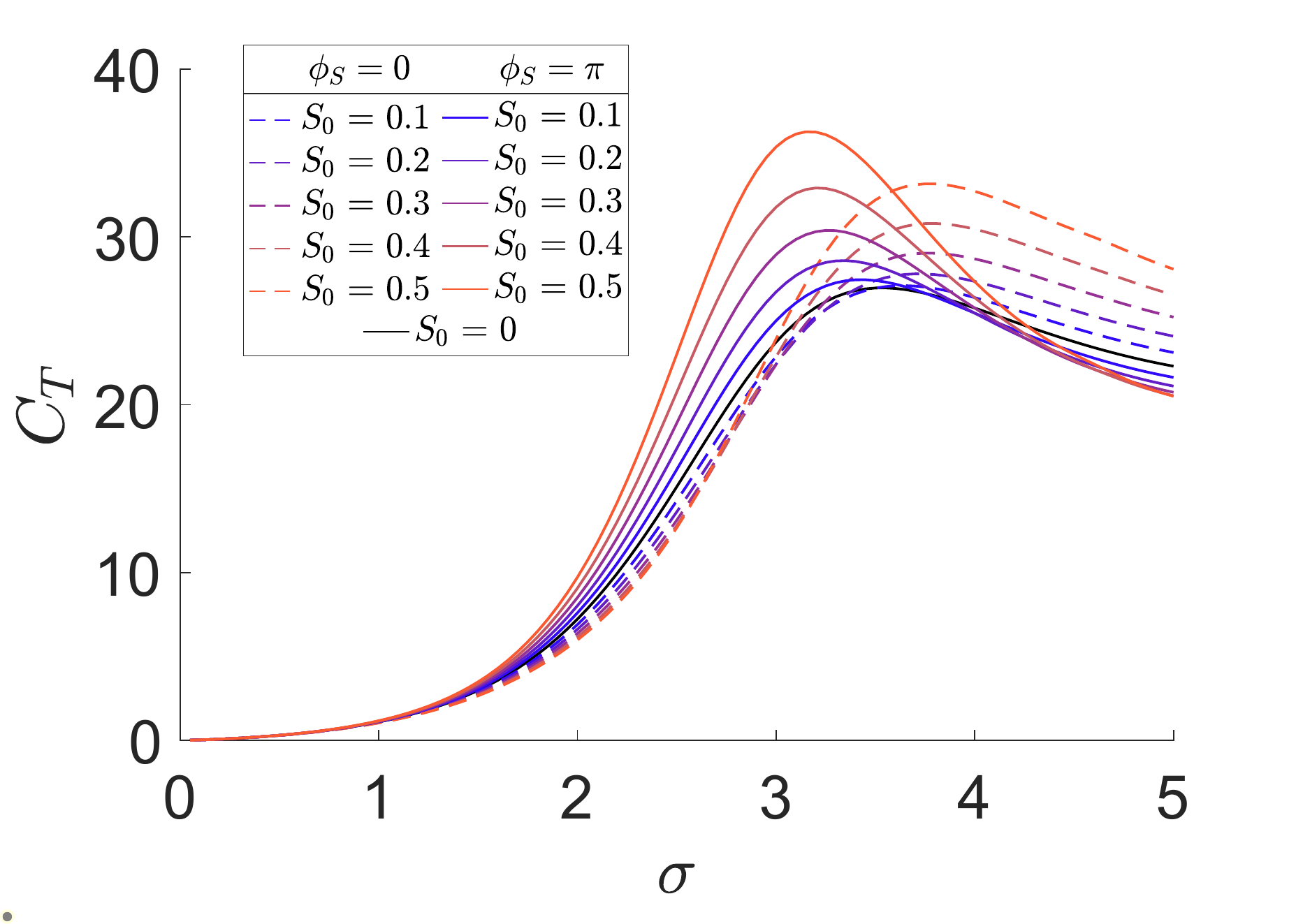}
  \caption{}
\end{subfigure}%
\begin{subfigure}{.5\textwidth}
  \centering
  \includegraphics[width=1\linewidth]{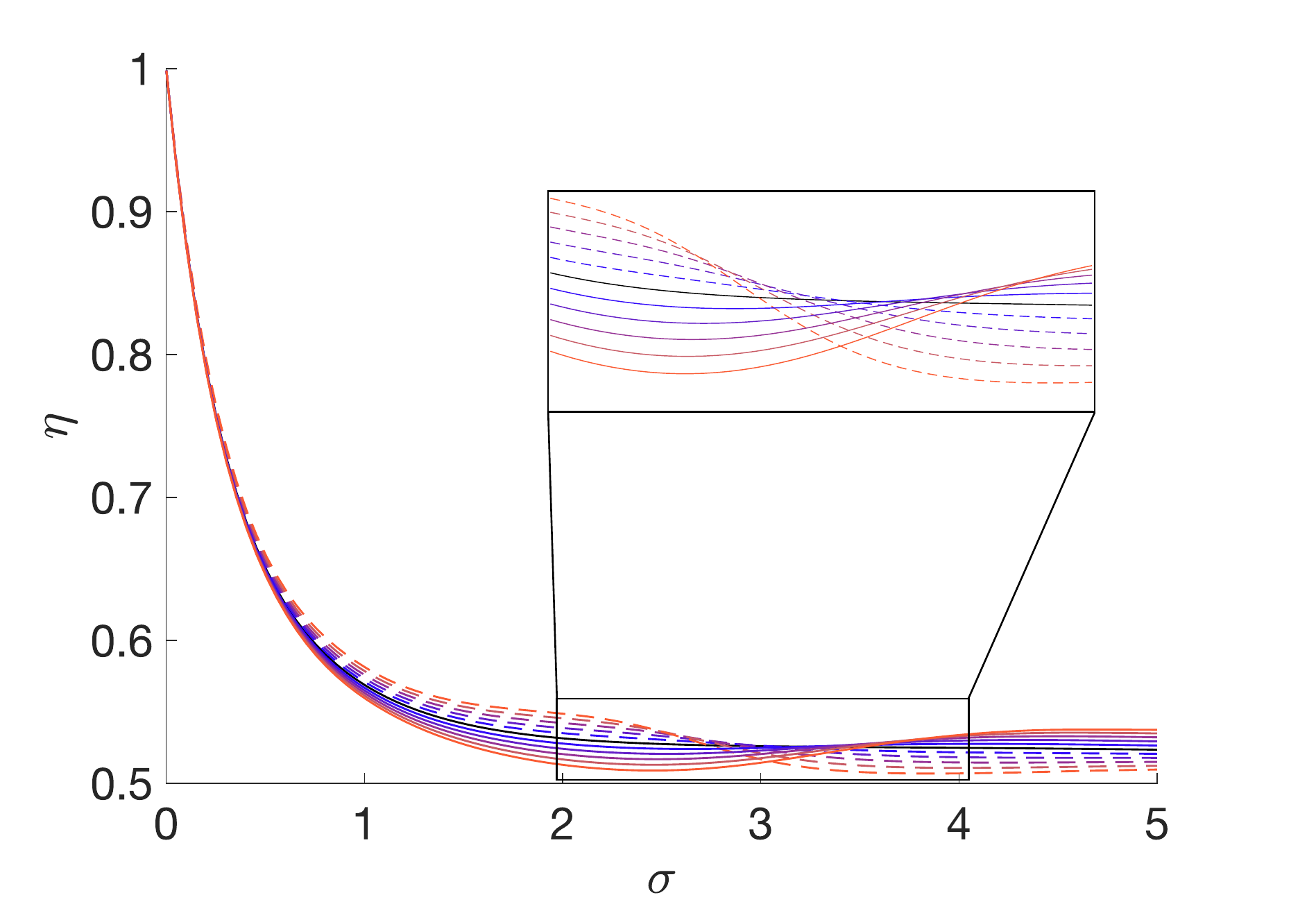}
  \caption{}
\end{subfigure}
    \caption{Thrust (left) and efficiency (right) for a heaving plate with gradually increasing stiffness oscillation amplitude. The parameters used are $\overline S = 20$, $h_0 = 1$, and $R = 0.01$.} 
      \label{fig:gradualCTincrease}
\end{figure}

In figure \ref{fig:gradualCTincrease}, we show how time-varying stiffness affects average thrust and efficiency. The frequency range is centered about the first resonant frequency of the plate with constant stiffness. Generally, adding periodic stiffness leads to a continuous and substantial increase in thrust as the stiffness oscillation amplitude increases, with up to a 35\% increase in thrust when $|S_0| = 0.5$. For $\phi_S = 0$, the resonance is shifted to higher frequencies, while for $\phi_S = \pi$ the resonance is shifted lower compared to the constant stiffness case. Past the resonant frequency, the $\phi_S = \pi$ case yields slightly lower thrusts than the baseline case.

For efficiency, we observe the opposite behavior (figure \ref{fig:gradualCTincrease}b). Performance benefits for ${\phi_S = 0}$ occur below the baseline resonant frequency, while for $\phi_S = \pi$ they occur above the baseline resonant frequency. For both phases of the stiffness oscillation, there are frequencies yielding greater efficiency than the constant-stiffness plate, as well as frequencies yielding lower efficiency. The effect of time-varying stiffness on the efficiency is much more mild than it is for thrust, however, as time-varying stiffness only changes the efficiency by $\Delta \eta \leq 0.05$. This suggests that time-varying stiffness may be a strategy to substantially increase thrust without much, if any, penalty in efficiency.


\begin{figure}
    \centering
    \begin{subfigure}{1\textwidth}
     \includegraphics[width = 0.9\linewidth]{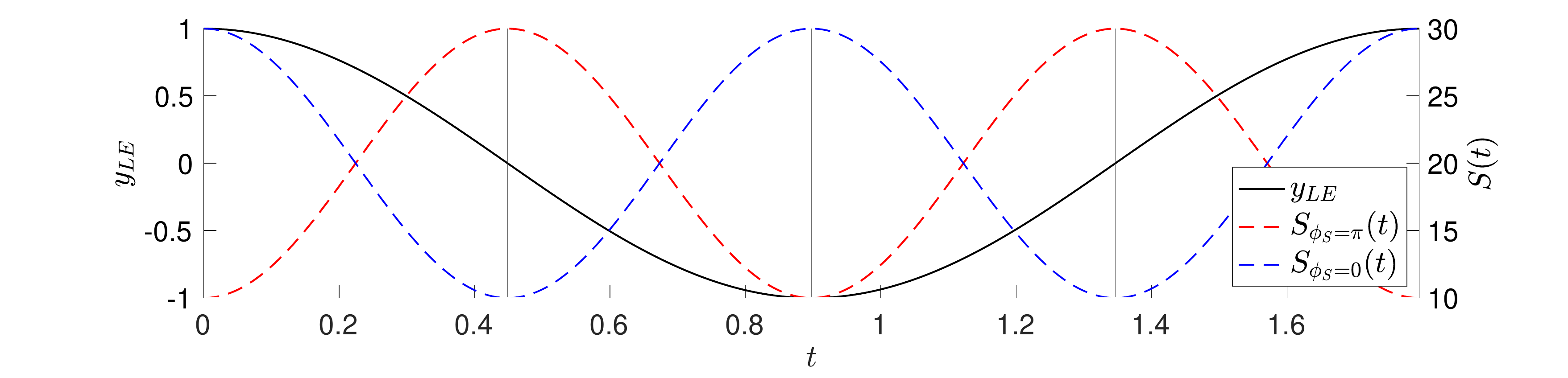}
     \caption{Kinematic input at the leading edge (solid black, left ordinate), and the stiffness distribution (blue and red dashed, right ordinate) for one stroke.}
    \end{subfigure} \\
    
    \begin{subfigure}{1\textwidth}
      \includegraphics[width = 0.9\linewidth]{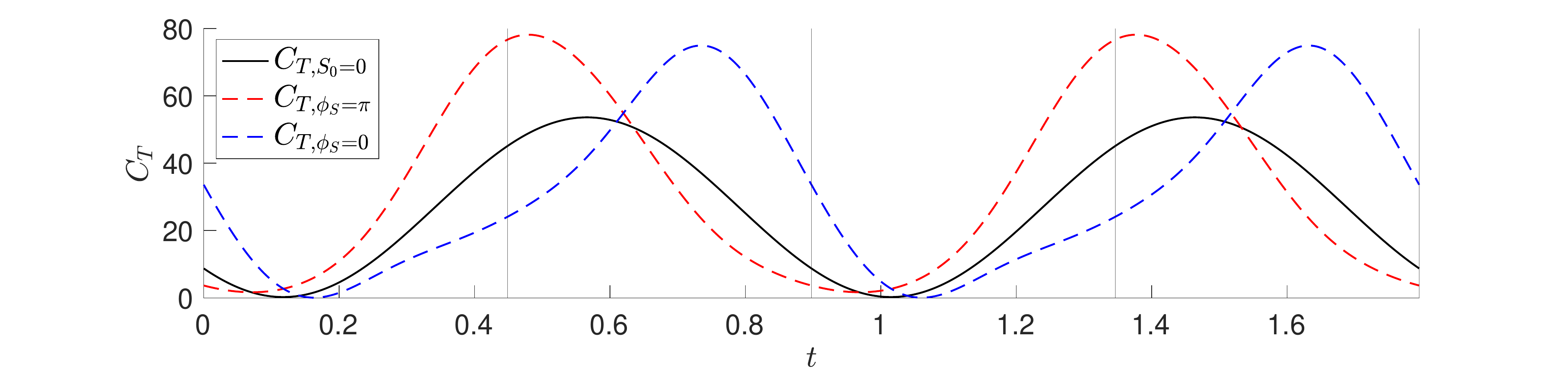}
      \caption{Thrust coefficient as a function of time.}
    \end{subfigure} \\
    \begin{subfigure}{1\textwidth}
      \includegraphics[width = 0.9\linewidth]{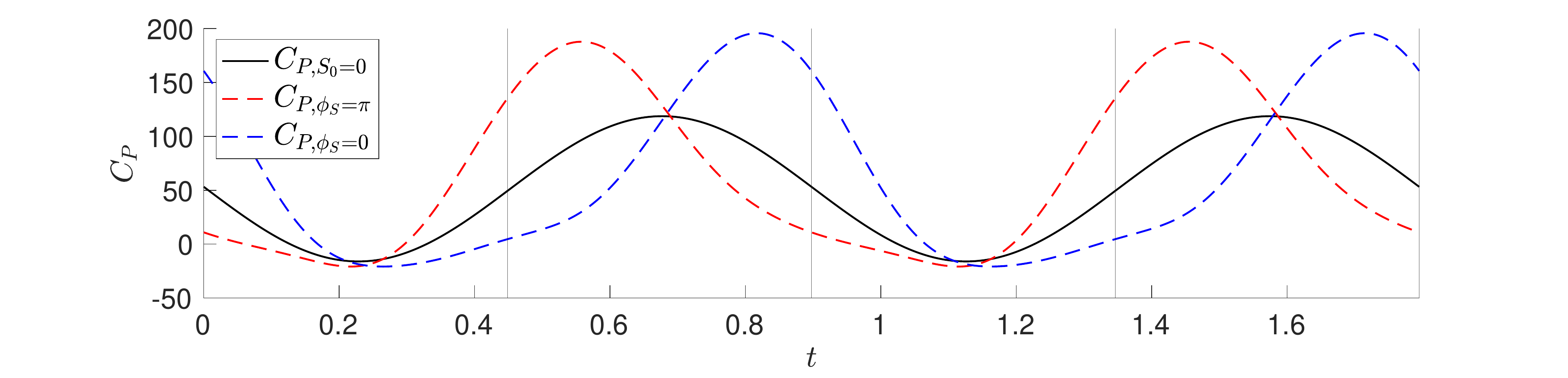}
      \caption{Power coefficient as a function of time. }
    \end{subfigure} \\
    \begin{subfigure}{1\textwidth}
      \includegraphics[width = 0.9\linewidth]{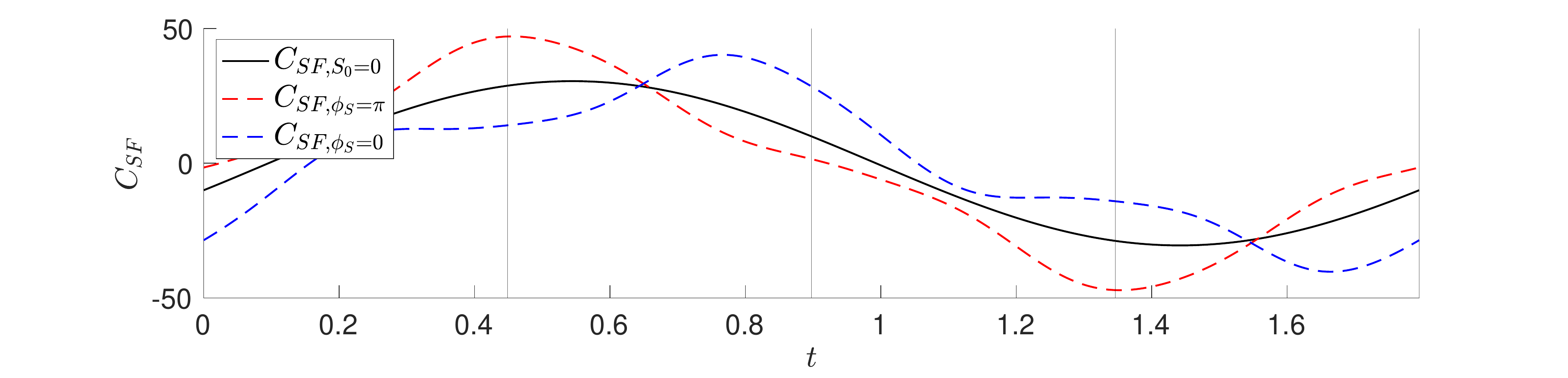}
      \caption{Side force coefficient as a function of time.}
    \end{subfigure} 
   \caption{Instantaneous thrust, power, and side force coefficients for a heaving plate. The parameters used are $\overline S = 20$, $h_0 = 1$, $R = 0.01$, $\sigma = 3.1$, and $|S_0| = 0.5$. Shown are a plate with constant stiffness (solid black), time-periodic stiffness in phase with the motion ($\phi_S = 0$; dashed blue), and time-periodic stiffness out of phase with the motion ($\phi_S = \pi$; dashed red).} 
    \label{fig:CF vs time}
\end{figure}

To better understand the effects of time-varying stiffness, we turn to the time histories of the performance characteristics over the course of an actuation cycle. Figure~\ref{fig:CF vs time} shows the instantaneous thrust, side force, and power coefficients over one period of the motion. For reference, we also plot the leading edge kinematics and plate stiffness. The plate with constant stiffness exhibits purely sinusoidal thrust, power, and side force since frequencies are uncoupled in the small-amplitude limit. Time-periodic stiffness, however, causes cross-frequency coupling, leading to non-sinusoidal behavior. Note that, because the stiffness oscillates at twice the frequency of motion, the side force remains symmetric, ensuring there is no mean side force (in a real system, this would lead to maneuvering).

The plate whose stiffness4 oscillates out of phase with the kinematics ($\phi_S = \pi$) produces the most thrust at the mid-stroke. During the turnaround the plate becomes the most flexible---this helps to relieve the power consumption which trends highest before the plate reverses direction. The opposite happens for the plate whose stiffness is in phase with the kinematics ($\phi_S = 0$). It produces the most thrust closer to the turnaround, and the power relief comes when the plate is most flexible at the maximum plunge velocity. For both cases, higher thrust trends towards the times of higher stiffness, and lower power consumption trends towards the times of lower stiffness. The phase of the stiffness oscillation dictates when in the cycle the plate is most stiff (promoting thrust) or most flexible (relieving power). Thus, by timing these stiffness changes at opportune times during the cycle (e.g., becoming flexible at a time when power is highest), the performance can be significantly enhanced. 

\begin{figure}
    \centering
    \begin{subfigure}{.5\textwidth}
  \centering
  \includegraphics[width=1\linewidth]{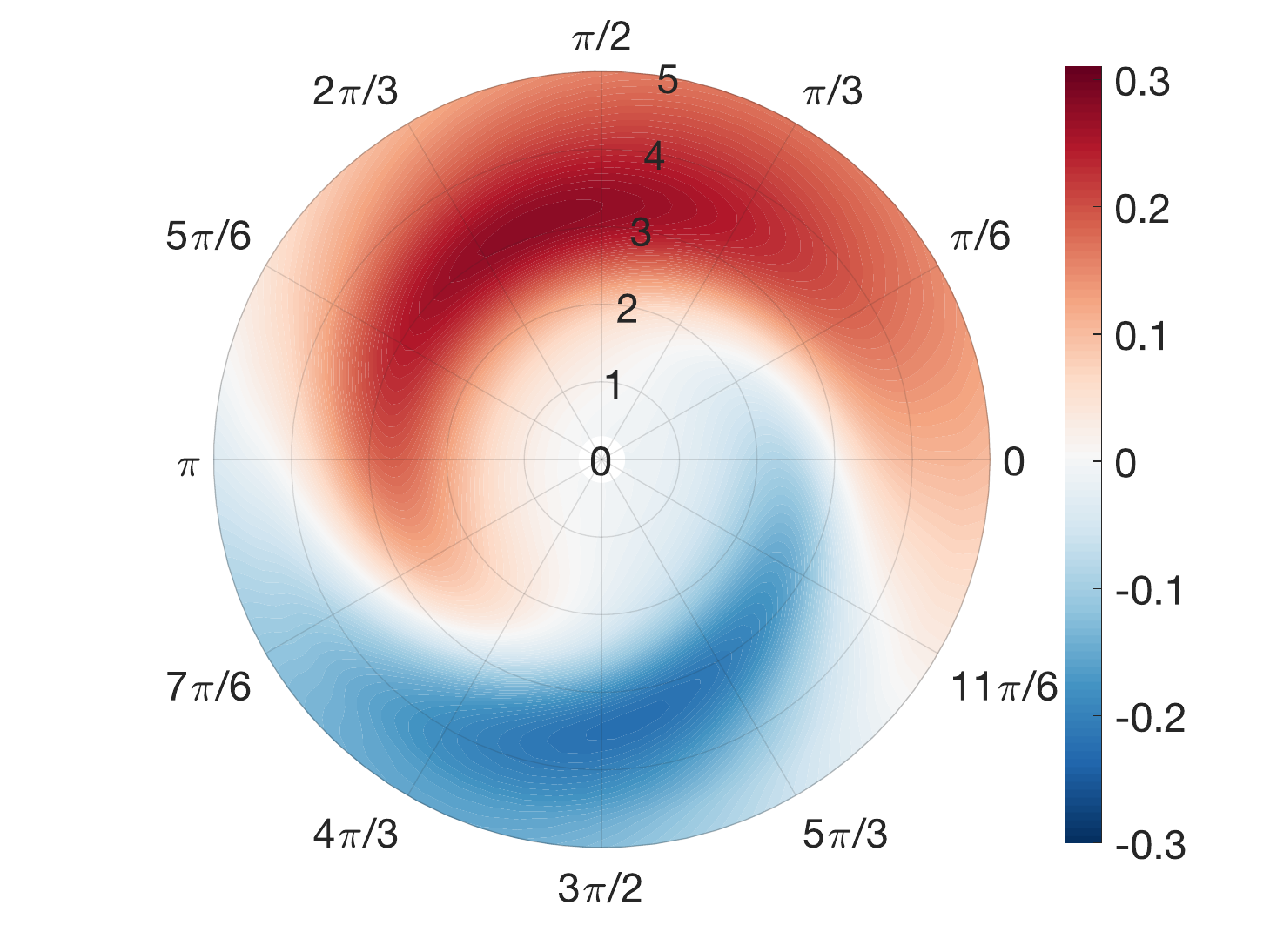}
  \caption{\small $\log{C_T/C_{T_{S_0=0}}}$}
\end{subfigure}%
\begin{subfigure}{.5\textwidth}
  \centering
  \includegraphics[width=1\linewidth]{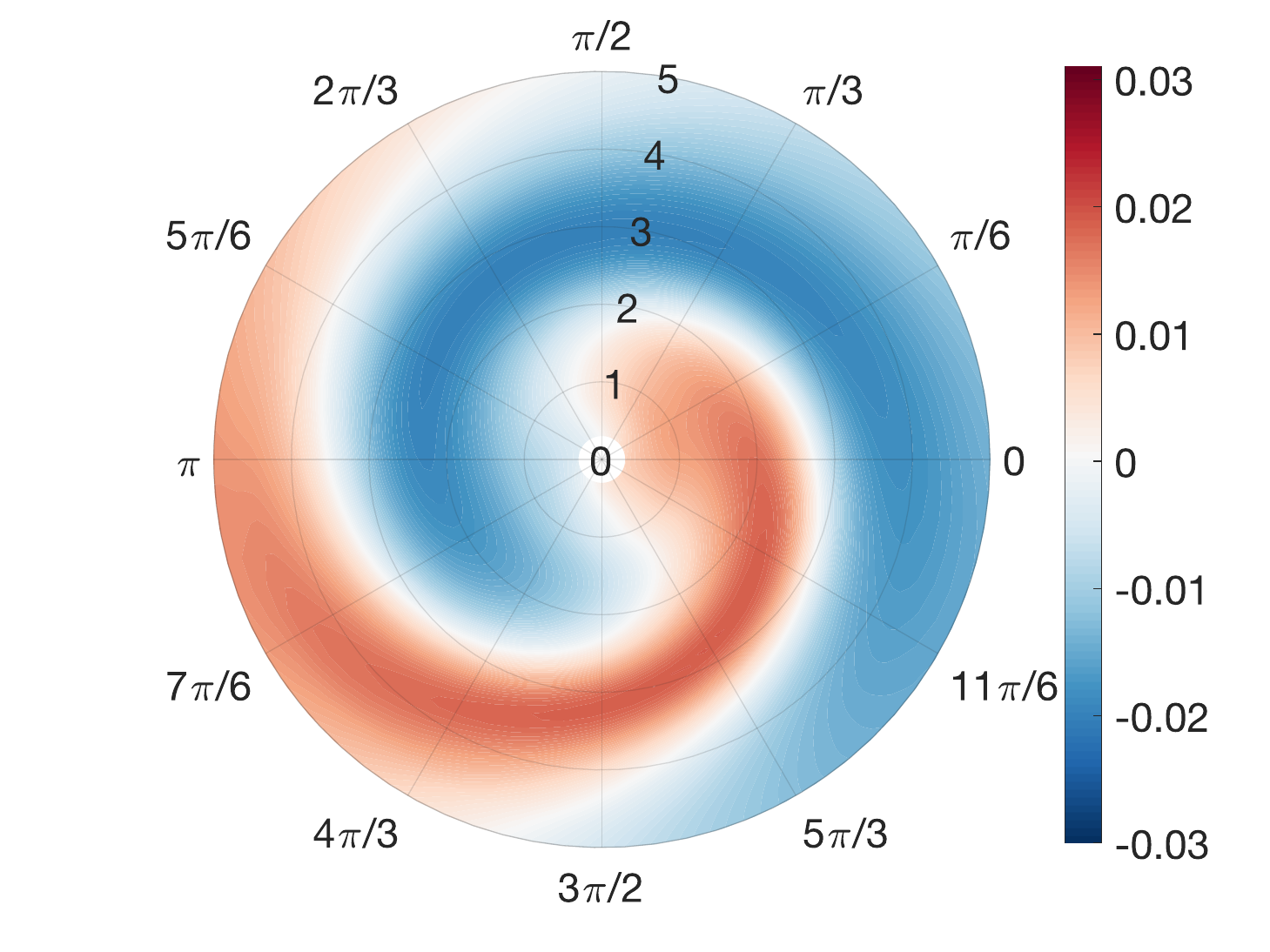}
   \caption{$\Delta \eta$}
\end{subfigure}
    \caption{ Impact of stiffness phase offset ($\phi_S$; azimuthal axis) and kinematic frequency ($\sigma$;~radial axis) on thrust and efficiency for a heaving plate, relative to the constant-stiffness case. The parameters used are $\overline S = 20$, $h_0 = 1$, $R = 0.01$, and $|S_0| = 0.5$.}
    \label{fig:impact of stiffness phase offset}
\end{figure}

It is clear that the timing of the stiffness oscillation is important. However, until this point, only two stiffness oscillation phases have been considered: in phase and out of phase with the kinematics. We investigate whether there is a particular phase that maximizes thrust or efficiency. In figure \ref{fig:impact of stiffness phase offset}, thrust and efficiency are plotted on a polar plot with frequency $\sigma$ on the radial axis and stiffness phase offset $\phi_s$ on the azimuthal axis. 
Both thrust and efficiency are shown relative to the constant-stiffness case. Time-varying stiffness increases thrust the most when $\phi_S$ is between $\pi/2$ and $2\pi/3$, and it increases efficiency the most near $\phi_s = 3\pi/2$. Again, the changes in efficiency are modest. It is important here to also highlight the importance of the kinematic frequency---for a given phase of the stiffness oscillation, there can be performance boosts and hindrances depending on the frequency of oscillation. 


\begin{figure}
\centering
\begin{subfigure}{.5\textwidth}
  \centering
  \includegraphics[width=1\linewidth]{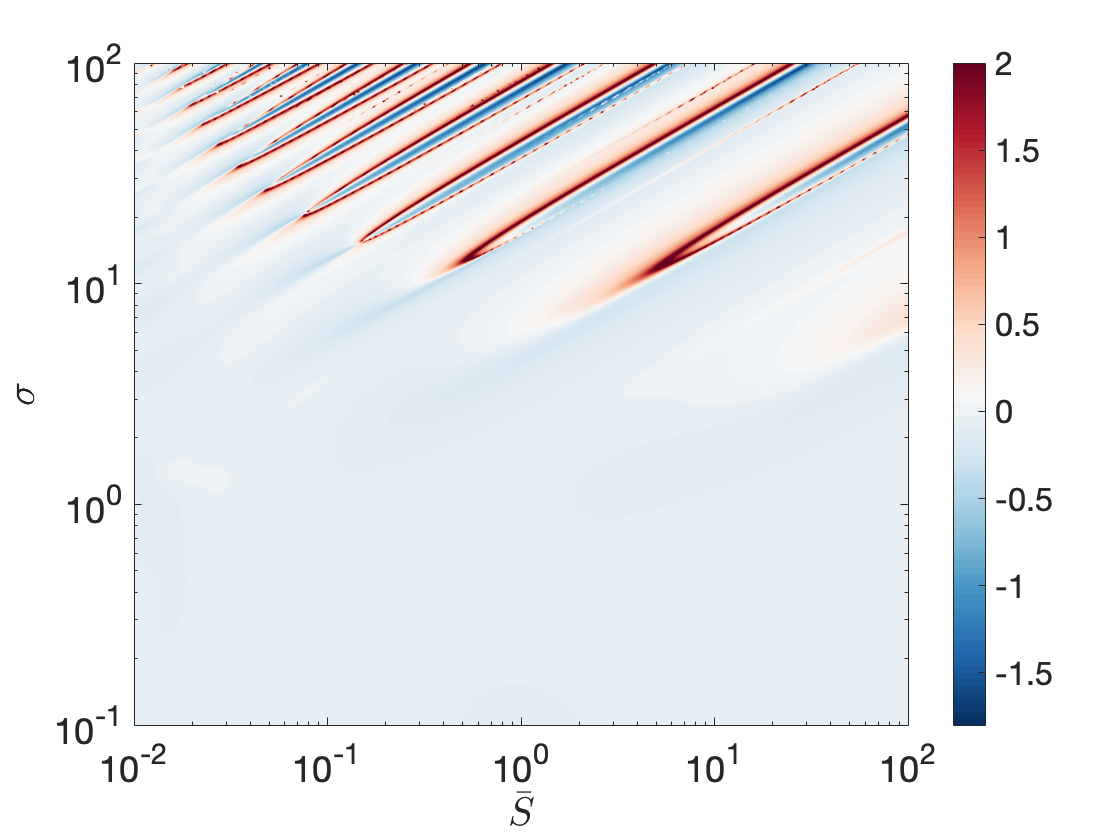}
  \caption{\small $\log{C_T/C_{T_{S_0=0}}}$ \\ $\phi_{S} = 0$}
  \label{fig:baselineContour_SUB1}
\end{subfigure}%
\begin{subfigure}{.5\textwidth}
  \centering
  \includegraphics[width=1\linewidth]{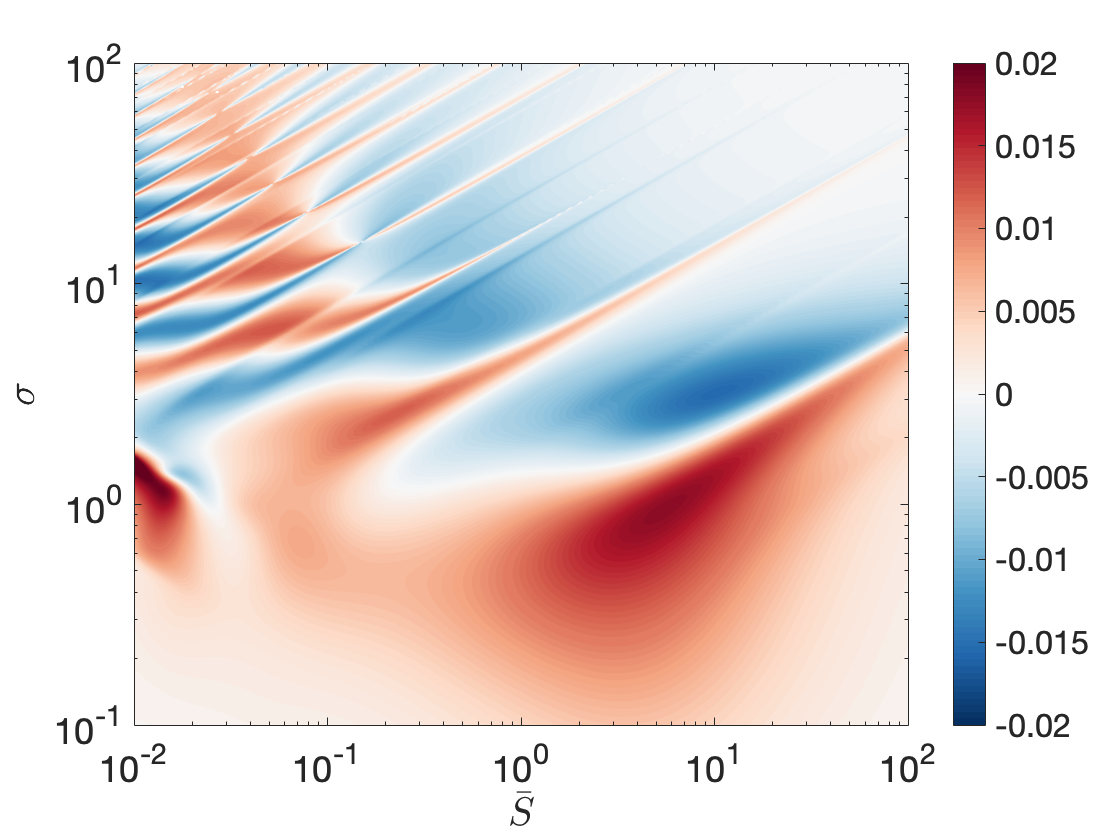}
  \caption{$\Delta \eta$ \\ $\phi_{S} = 0$ }
  \label{fig:baselineContour_SUB2}
\end{subfigure}%
\vskip\baselineskip
\begin{subfigure}{.5\textwidth}
  \centering
  \includegraphics[width=1\linewidth]{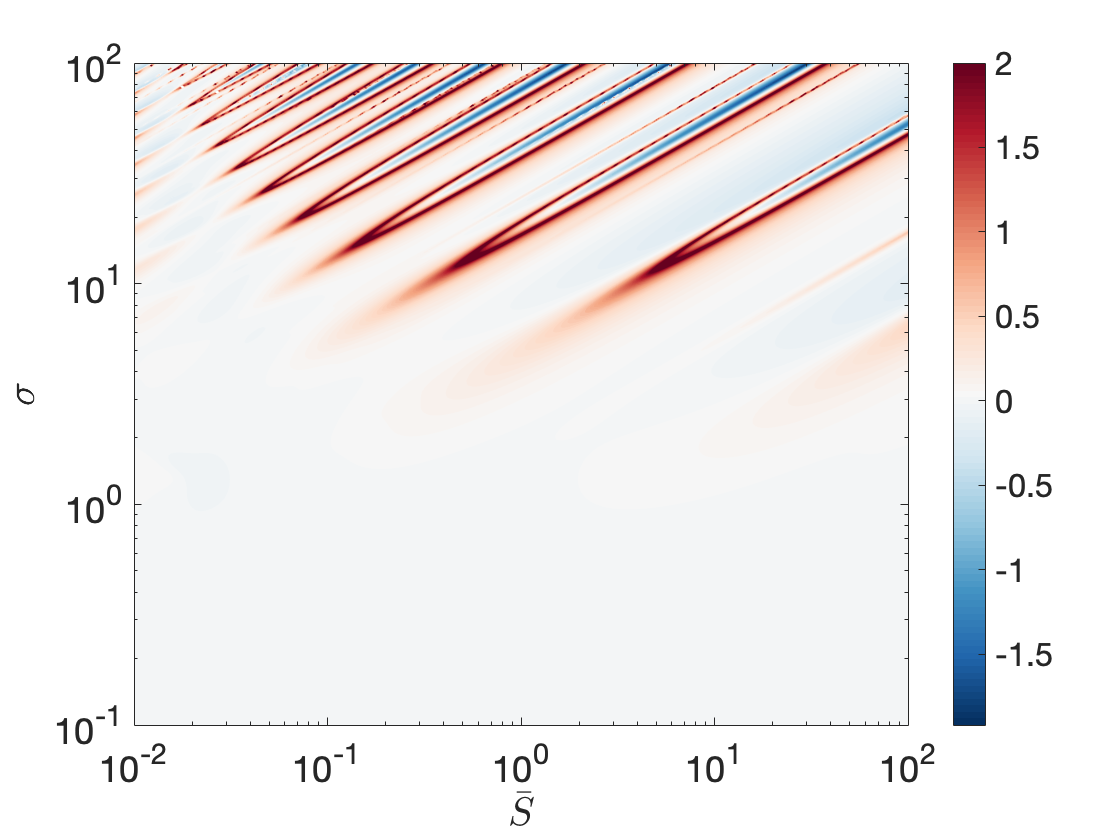}
  \caption{\small $\log{C_T/C_{T_{S_0=0}}}$ \\ $\phi_{S} = \pi$}
  \label{fig:baselineContour_SUB3}
\end{subfigure}%
\begin{subfigure}{.5\textwidth}
  \centering
  \includegraphics[width=1\linewidth]{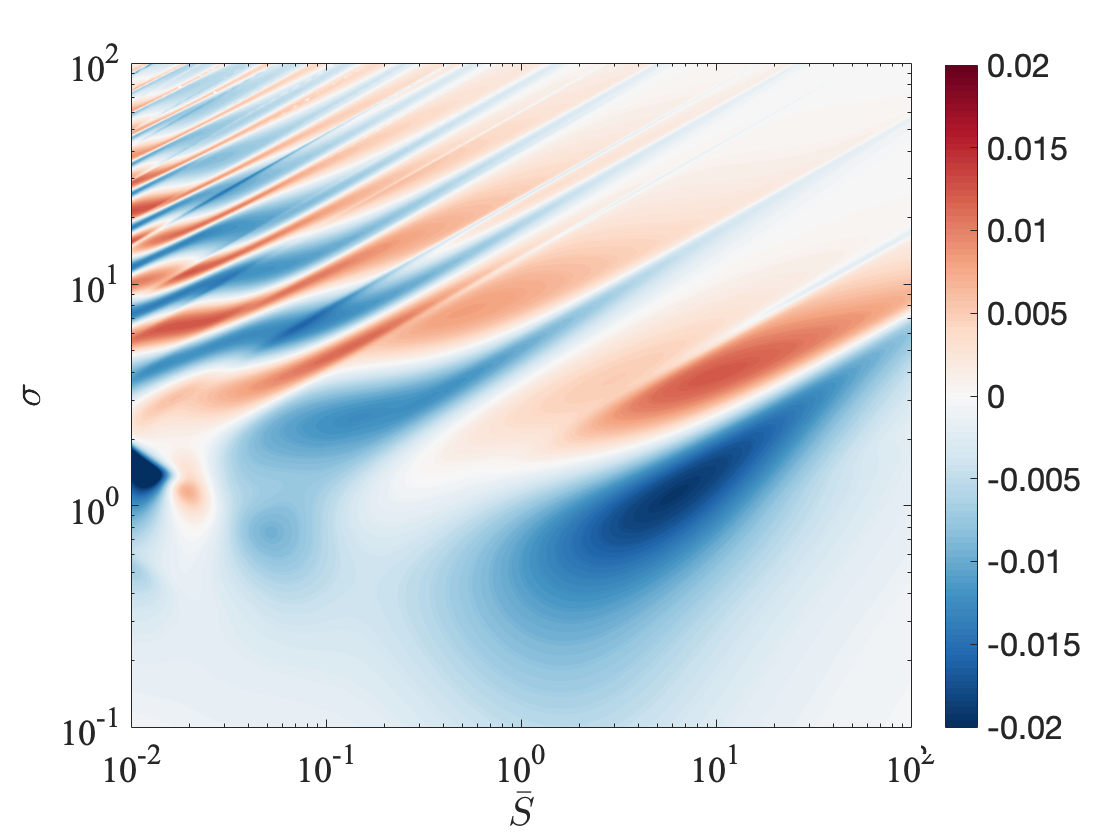}
  \caption{$\Delta \eta$ \\ $\phi_{S} = \pi$ }
  \label{fig:baselineContour_SUB4}
\end{subfigure}
\caption{Impact of time-varying stiffness on the thrust (left) and efficiency (right), relative to the constant-stiffness case, for stiffness oscillations that are in phase with the motion (top), and 180$^\circ$ out of phase with the motion (bottom). The parameters used are $h_0 = 1$, $R = 0.01$, and $|S_0| = 0.5$.}
\label{fig:baselineContour}
\end{figure}

We now explore a much wider range of mean plate stiffnesses and frequencies encompassing higher-order resonant frequencies. The change in swimming performance due to time-periodic stiffness, with respect to the constant-stiffness case, is shown in figure \ref{fig:baselineContour} for $\phi_S = 0$ and $\pi$. The oscillation in stiffness modifies the resonant frequencies from those of the constant-stiffness plate, as we saw in figure~\ref{fig:gradualCTincrease}, leading to sharp bands of increased and decreased thrust; these appear as adjacent narrow red and blue strips in figure~\ref{fig:baselineContour}. In a linear time-invariant system, resonant frequencies are related to the imaginary parts of the eigenvalues of the system, whereas in linear time-periodic systems, they are related to the imaginary parts of the Floquet exponents \citep{wereley1990analysis}. In general, the two are different, leading to the modified resonant frequencies that we observe. Away from resonance, thrust is modified little by the oscillation in stiffness. The changes in efficiency are much broader over the stiffness-frequency plane and have features that align with resonant frequencies; they are, however, very mild. Generally, thrust increases where efficiency decreases, but increasing $|S_0|$ can greatly increase the thrust production with minimal impact on efficiency. {Note that we are using the small amplitude deflection and slope assumption while exploring resonance up to the second mode. One might assume that our assumption is invalid, due to the high deflections and slopes at the second resonance or higher. On the other hand, the kinematics of the plate scale linearly with the leading edge input. Therefore if we notice a deflection that would violate our assumption, we can just scale down our inputs until the deflection and slope satisfies the linear theory again.} 

\begin{figure}
    \centering
    \begin{subfigure}{0.5\textwidth}
    \includegraphics[width=1\linewidth]{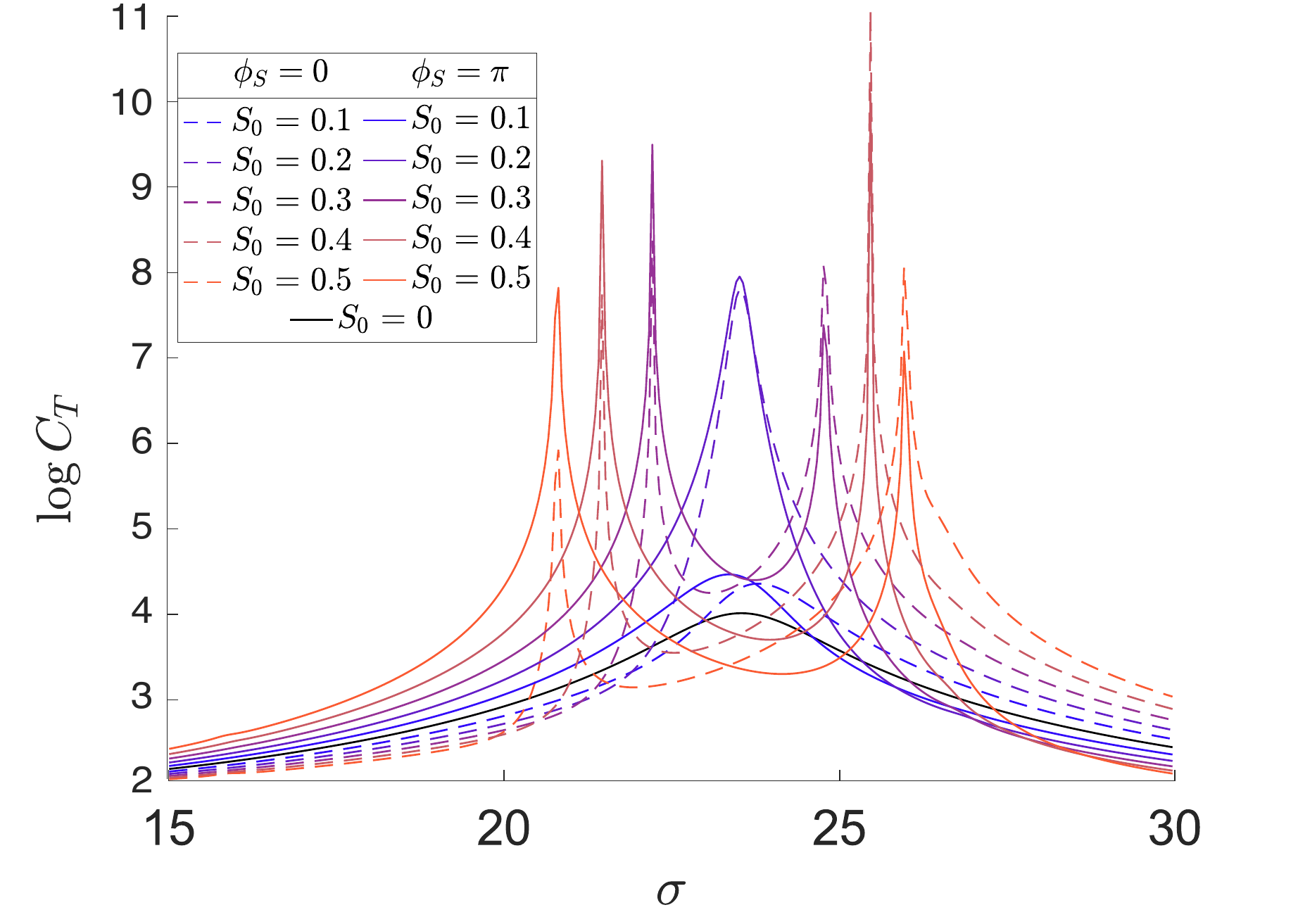}
    \caption{}
    \end{subfigure}%
    \begin{subfigure}{0.5\textwidth}
      \includegraphics[width = 1\textwidth]{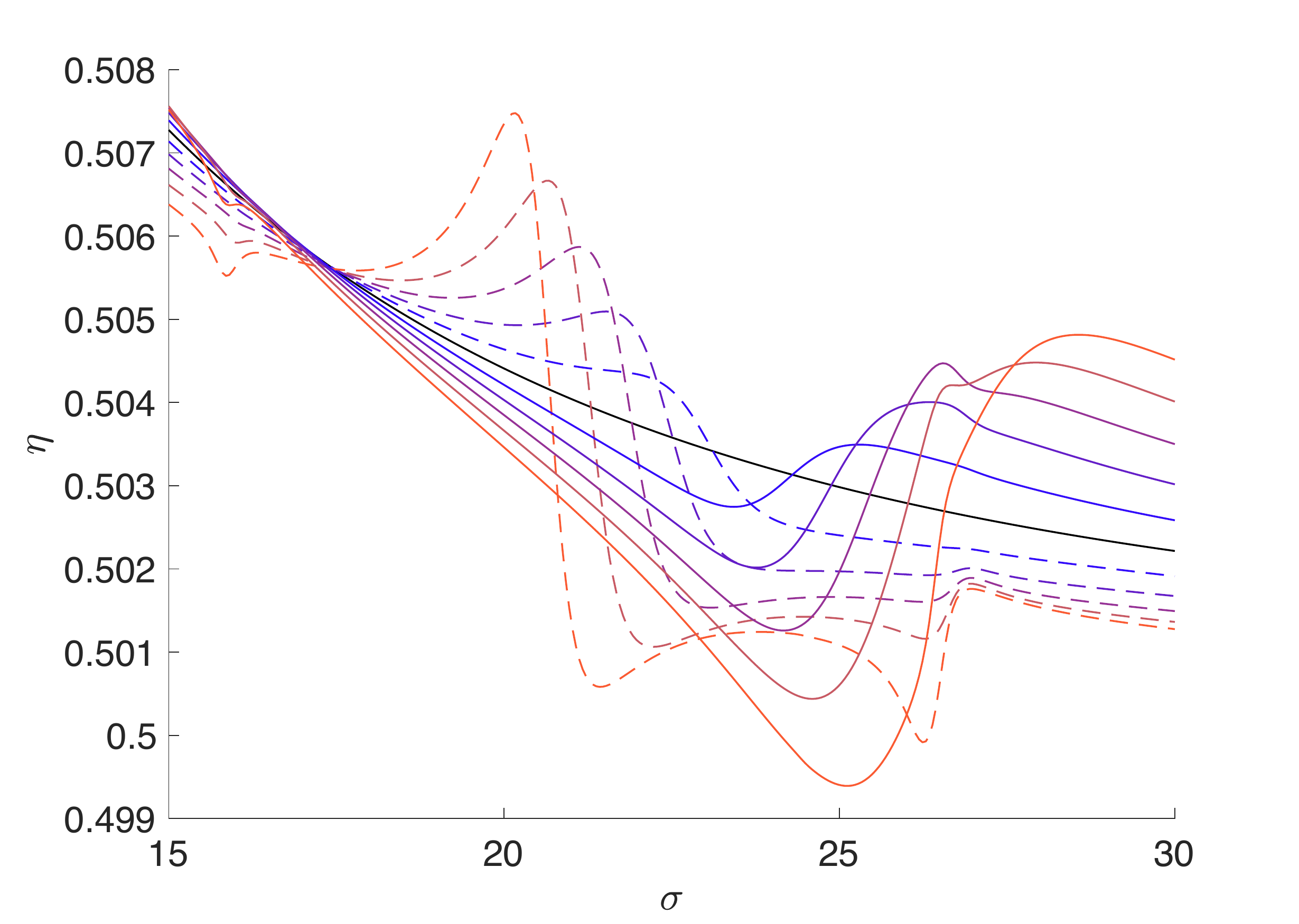}
      \caption{}
    \end{subfigure}
    \caption{$\log C_T$ (left) and efficiency (right) for a heaving plate with gradually increasing stiffness oscillation amplitude. As stiffness oscillation amplitude $|S_0|$ increases, a single resonant peak bifurcates into two. The parameters used are $\overline S = 20$, $h_0 = 1$, and $R = 0.01$.}
    \label{fig:birfucation peaks}
\end{figure}

Perhaps the most interesting features in figure \ref{fig:baselineContour} are the hollowed peaks in the thrust, resembling the eye of a needle. They are more prominent at higher frequencies. To clarify this behavior, in figure \ref{fig:birfucation peaks} we show the thrust and efficiency along a slice in the stiffness-frequency plane, taking $\overline{S} = 20$ and centering the frequency range about the second resonant frequency (this is the same parameter case as shown in figure \ref{fig:gradualCTincrease}). As the stiffness oscillation amplitude is gradually increased, there is a critical value of $|S_0|$ at which a single resonant peak in thrust bifurcates into two sharp peaks centered about the baseline resonant frequency, with the distance between the peaks increasing as $|S_0|$ increases. The bifurcated peaks are very sharp, indicative of natural frequencies with very little damping. {Low damping leads to more energy input through the time-periodic stiffness then is dissipated through the wake and material damping, which leads to a net energy increase of the system ($\frac{d E}{dt} >0, \forall t > 0$). This causes small perturbations in the kinematics to grow unbounded.} To investigate this possibility, we perform a Floquet analysis. 

\begin{figure}
    \centering
    \includegraphics[width = 0.75\textwidth]{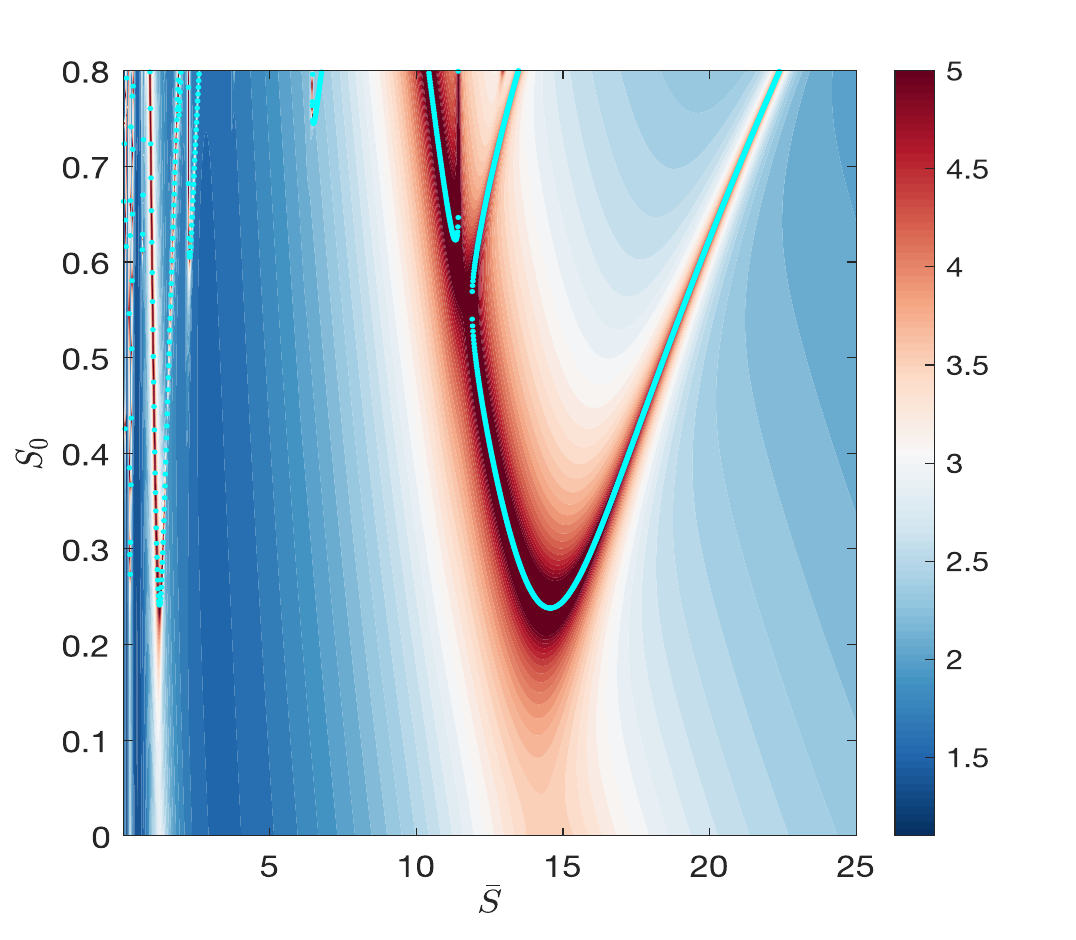}
    \caption{Neutral stability curves overlaid on contours of $\log C_T$. The parameters used are $R = 0.01$ to calculate the neutral stability curves, and $h_0 = 1, \sigma = 20,$ and $\phi_S = 0$ to calculate the thrust values.}
    \label{fig:floquetPlate}
\end{figure}

Representative results from the Floquet analysis are shown in figure~\ref{fig:floquetPlate}. There, we have plotted the neutral curves in the $\overline S$-$|S_0|$ plane on top of contours of $\log C_T$. When the stiffness oscillation amplitude $|S_0|$ is below the neutral curves, the system is stable. Conversely, the system is unstable when $|S_0|$ is above the neutral curves. We see the emergence of tongues of instability, which are characteristic features of parametrically excited systems such as Mathieu's equation \citep{nayfeh1995nonlinear} and Faraday waves \citep{faraday1831peculiar, miles1990parametrically}. Because the system is damped (due to energy lost to the wake via a thin sheet of vorticity), the instability emerges at a non-zero value of $|S_0|$. For a fixed value of $|S_0|$, as we vary the mean stiffness $\overline S$, we cross into and then out of the unstable region. At the boundaries, the thrust exhibits sharp peaks, which is explained by the theory of forced linear time-periodic systems \citep{wereley1990analysis}. This explains the double resonant peaks that we observed in figures~\ref{fig:baselineContour} and~\ref{fig:birfucation peaks}. (Although we vary $\sigma$ at a fixed value of $\overline S$ in figure~\ref{fig:birfucation peaks}, we can see in figure~\ref{fig:baselineContour} that we encounter double resonant peaks whether we vary $\overline S$ at a fixed value of $\sigma$, or we vary $\sigma$ at a fixed value of $\overline S$; physically, the origin of the double resonant peaks is the same.) No physical significance should be given to the thrust in the unstable region between the double resonant peaks since the thrust was calculated under the assumption of a stable system. We have verified for conditions other than those used to generate figure~\ref{fig:floquetPlate} that a splitting of one resonant peak into two coincides with the emergence of an instability; we conjecture that every region between double resonant peaks in figures~\ref{fig:baselineContour} and~\ref{fig:birfucation peaks} is actually unstable. While instability usually has a negative connotation in engineering applications, these unstable regions could potentially lead to greatly enhanced propulsive performance since an instability would produce large deflections from small actuation. Our linear method cannot capture the saturation of the instabilities, but we believe that exploring the unstable regions via experiments or nonlinear simulations is a promising direction. We believe that \cite{doi:10.1063/5.0027927} does not see these instabilities due to the use of the nonlinear beam equation, which would delay the emergence of the instabilities to higher resonant frequences.

\subsection{Pitch-only and pitch-and-heave motions}

We now consider how pitching the leading edge changes the results. As stated previously, pitching and heaving are fundamentally different in their thrust generation mechanism \citep{floryan2017scaling,van2020bioinspired}, with the former utilizing added-mass forces and the latter using lift-based or circulatory forces. Additionally, combined pitching and heaving motions are the most biologically relevant and also the best-performing \citep{VanBurenAIAA, doi:10.1146/annurev-fluid-122109-160648}. In this section, we study two cases: (1) purely pitching, and (2) combined pitching and heaving with pitch lagging heave by $\pi / 2$, the most `fish-like' motion that is also generally the most efficient \citep{VanBurenAIAA, doi:10.1146/annurev.fluid.32.1.33}. 

\begin{figure}
    \centering
    \begin{subfigure}{1\textwidth}
          \includegraphics[width = 1\textwidth]{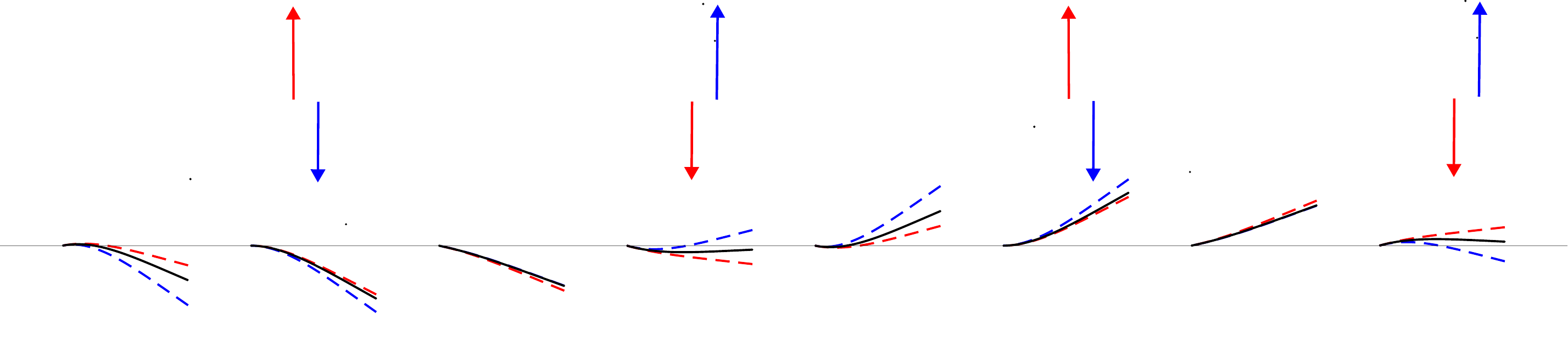}
          \caption{}
    \end{subfigure} \\
    \begin{subfigure}{1\textwidth}
      \includegraphics[width = 1\textwidth]{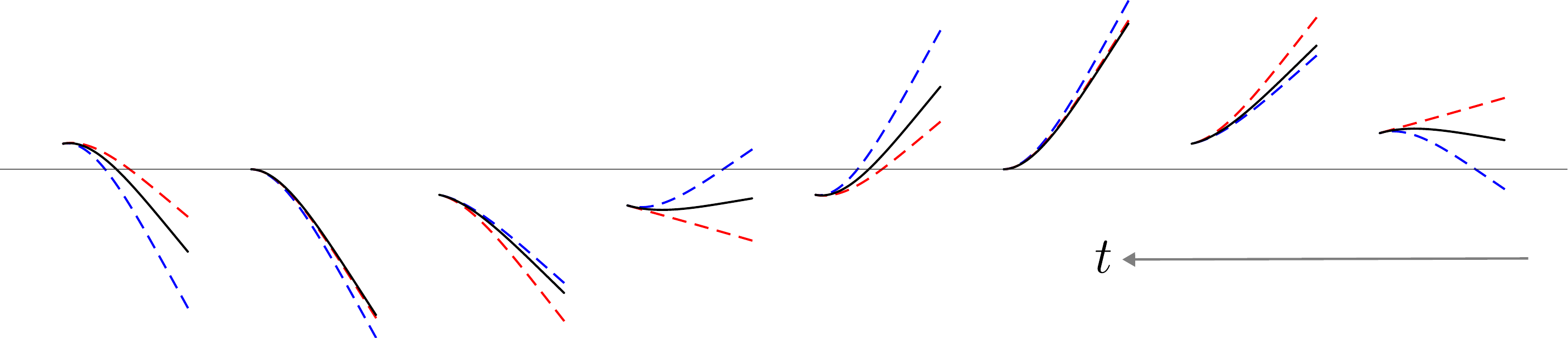}
      \caption{}
    \end{subfigure}
   \caption{Analogous to figure~\ref{fig:baselineHeaveKinematics}, but for a pitching plate (top) and a pitching and heaving plate (bottom). Shown are a plate with constant stiffness (solid black), time-periodic stiffness in phase with the motion ($\phi_S = 0$; dashed blue), and time-periodic stiffness out of phase with the motion ($\phi_S = \pi$; dashed red). The parameters used are $\overline S = 20$, $h_0 = 1$, $\theta_0 = -0.5j$, $R = 0.01$, and $|S_0| = 0.5$.}
    \label{fig:baselinePitchKinematics}
\end{figure}

Figure \ref{fig:baselinePitchKinematics} shows the kinematics for both cases during one cycle of motion. The kinematics are qualitatively the same as for the purely heaving case in figure \ref{fig:baselineHeaveKinematics}. Adding an oscillatory component to the stiffness causes a lead/lag effect on the kinematics which dictates at what point in the stroke thrust is generated. When the stiffness is in phase with the motion, rapid trailing edge motion occurs near the turnaround where the plate transitions from flexible to stiff, whereas when the stiffness is out of phase with the motion, the rapid trailing edge motion occurs through the midpoint of the cycle. Even for the pitching and heaving case---which specifically reduces the side-force on the plate \citep{VanBurenAIAA}---we see the stiffness oscillations have similar impact in both magnitude and timing. 

\begin{figure}
    \centering
    \begin{subfigure}{.5\textwidth}
  \centering
  \includegraphics[width=1\linewidth]{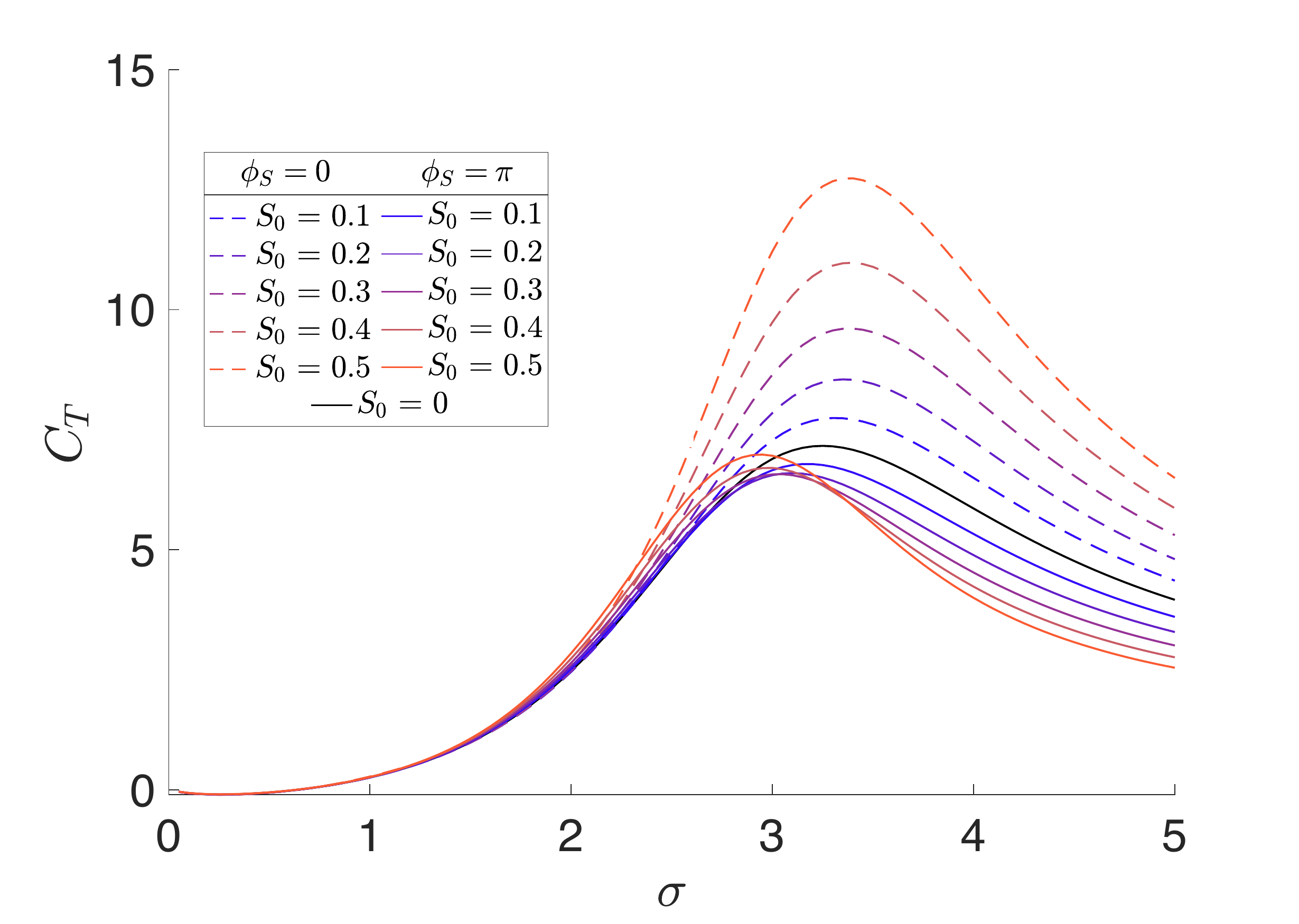}
  \caption{$C_T$ \\
  $h = 0, \theta = 0.5$}
\end{subfigure}%
\begin{subfigure}{.5\textwidth}
  \centering
  \includegraphics[width=1\linewidth]{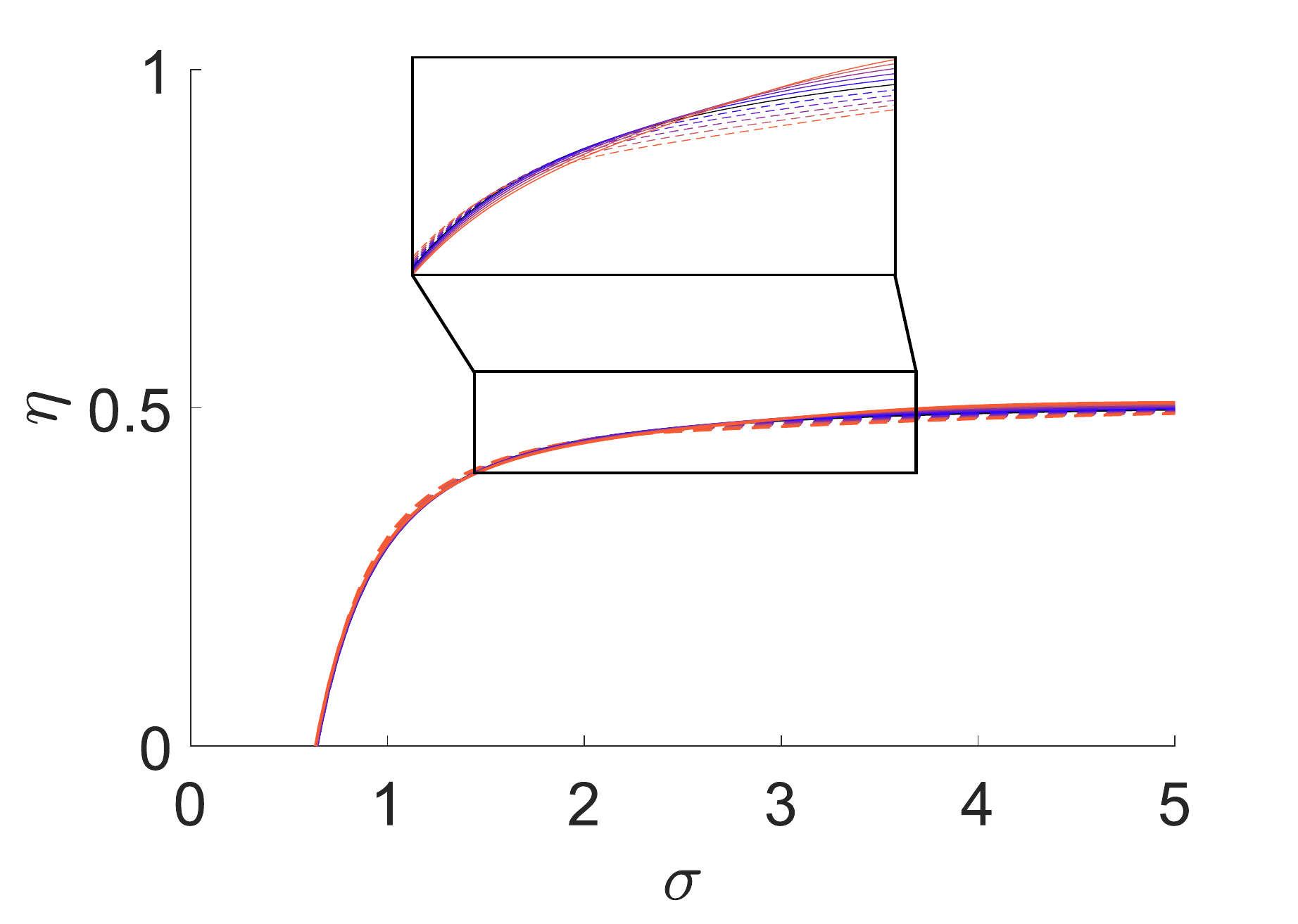}
  \caption{$\Delta \eta$ \\
  $h = 0, \theta = 0.5$}
\end{subfigure} \\
\begin{subfigure}{.5\textwidth}
  \centering
  \includegraphics[width=1\linewidth]{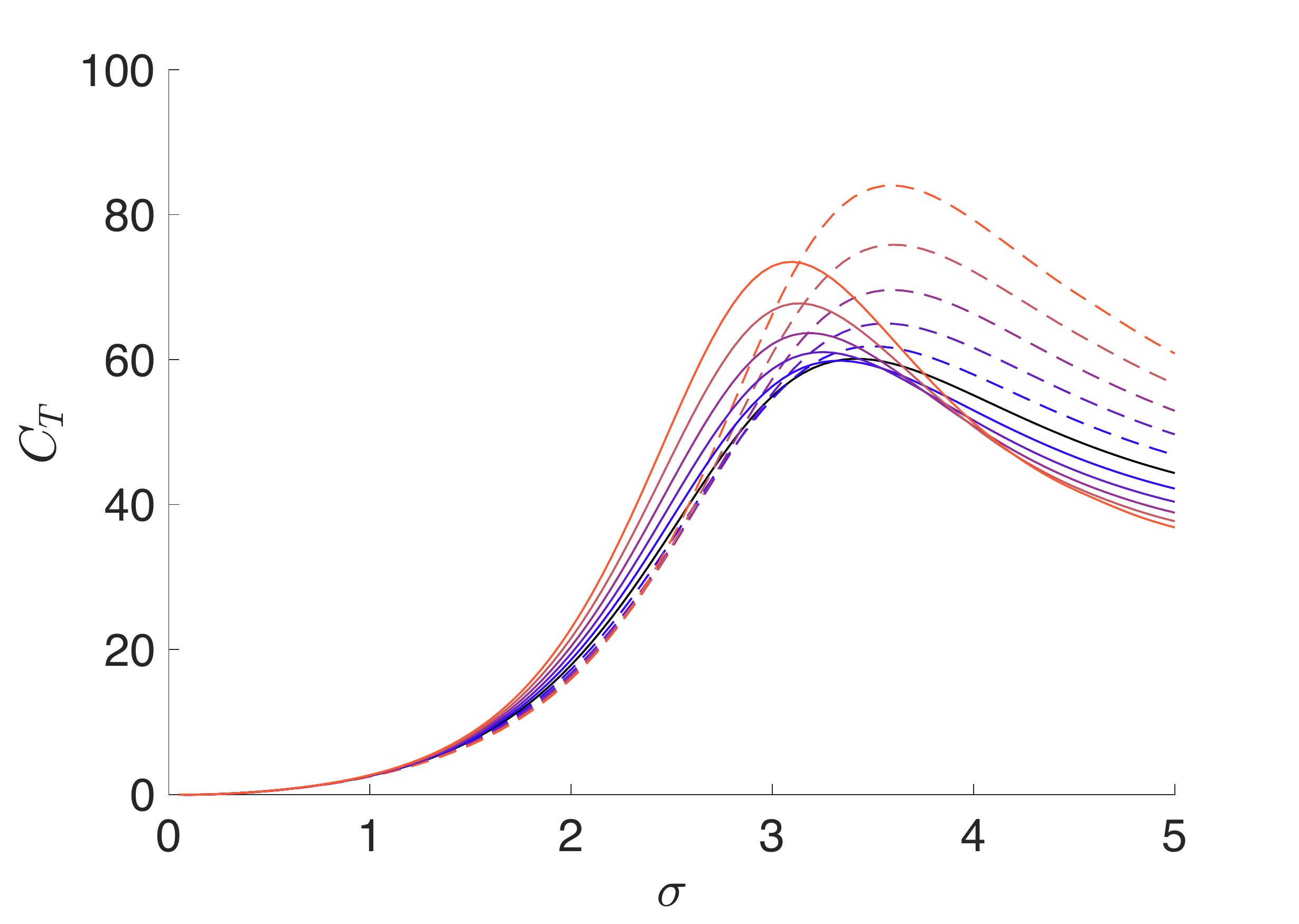}
  \caption{$C_T$ \\
  $h = 1, \theta = 0.5$}
\end{subfigure}%
\begin{subfigure}{.5\textwidth}
  \centering
  \includegraphics[width=1\linewidth]{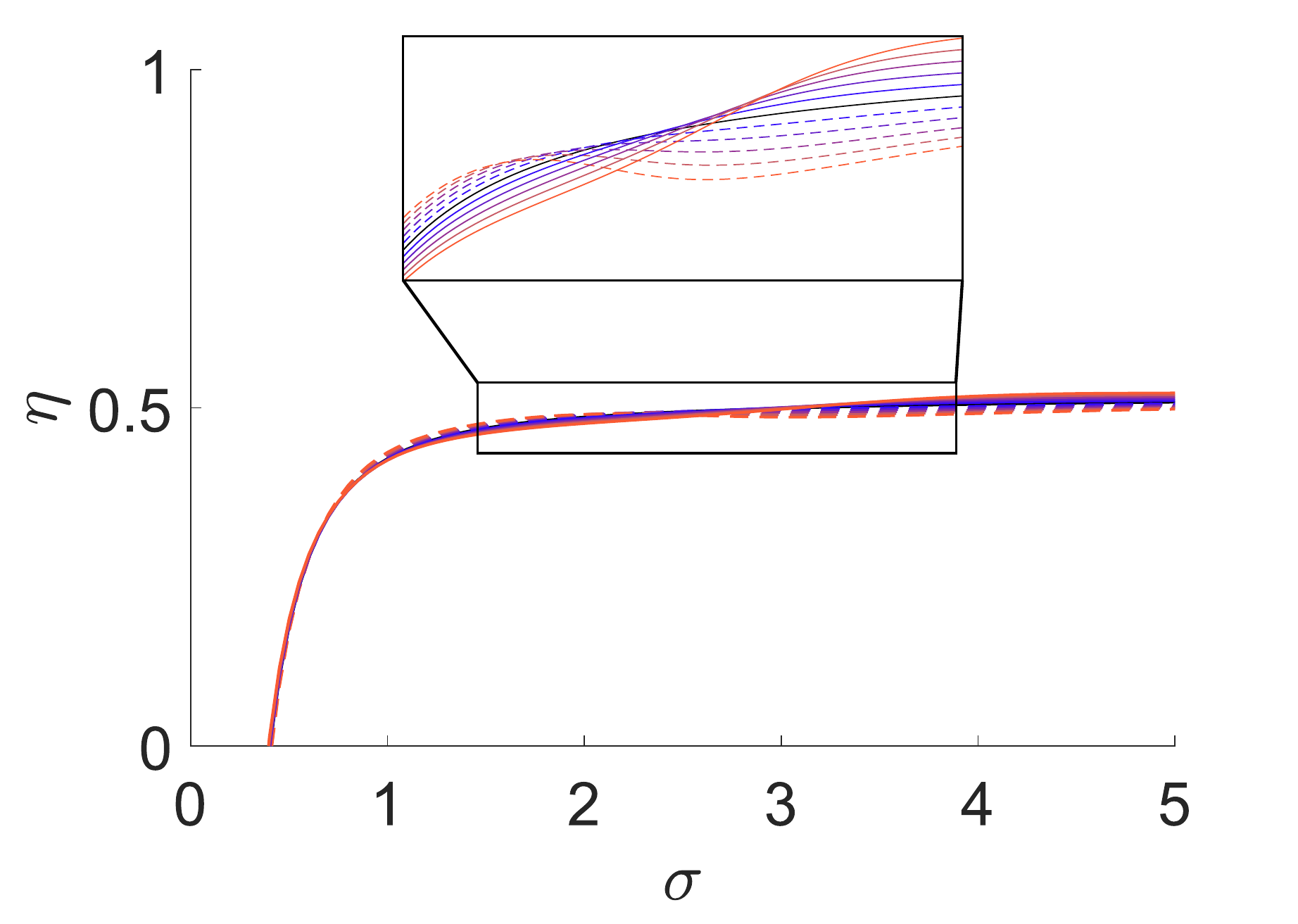}
  \caption{$\Delta \eta$ \\
  $h = 1, \theta = 0.5$}
\end{subfigure}
    \caption{Thrust and efficiency of a purely pitching plate (top) and a pitching and heaving plate (bottom) as a function of driving frequency $\sigma$. The parameters used are $\overline S = 20$, $h_0 = 1$, $\theta_0 = -0.5j$, and $R = 0.01$.}
    \label{fig:pitched CT/eta}
\end{figure}

Figure \ref{fig:pitched CT/eta} shows the thrust and efficiency for the two cases in a frequency range centered about the first resonant frequency of the plate with constant stiffness. For both the pitch and pitch-and-heave cases, we see that the oscillating stiffness has a strong impact on the peak thrust near resonance. The efficiency is less impacted by the stiffness oscillation and switches from being greater than the efficiency of a constant-stiffness plate to less than it across the resonant frequency---depending on the phase of the stiffness oscillation---which is similar to the behavior of the purely heaving plate in figure \ref{fig:gradualCTincrease}. For the purely pitching plate, thrust is much more impacted by the oscillating stiffness when it is out of phase with the kinematics, $\phi_S = 0$. For pitching, the blue plate lags behind the red and black plate at the turnaround, and becomes the least stiff at this moment. The blue plate accelerates the fastest between frames 2 and 4 in figure \ref{fig:baselinePitchKinematics}, which corresponds to the fastest angular velocity at the leading edge. The blue plate had the highest trailing edge amplitude, because the phase difference between the leading edge input and the trailing edge deflection and its lower stiffness allows the blue plate to reach a higher trailing edge amplitude before feeling the effects of the acceleration at the leading edge. The blue plate then becomes stiffer as the effect of the acceleration at the leading edge reaches the trailing edge. This combination of high acceleration, high trailing edge amplitude, and high stiffness creates high thrust. The red plate's stiffness distribution is misaligned with the phase difference caused by pure pitching, and therefore reaches a lower max trailing edge amplitude, and benefit less from the high acceleration of the leading edge between frames 2 and 4 in figure \ref{fig:baselinePitchKinematics}. For a better visualization of the plate dynamics and how they tie into the performance enhancements, refer to the online supplemental movies where we show the heave, pitch, and heave plus pitch kinematics.


\begin{figure}
    \centering
    \begin{subfigure}{.5\textwidth}
  \centering
  \includegraphics[width=1\linewidth]{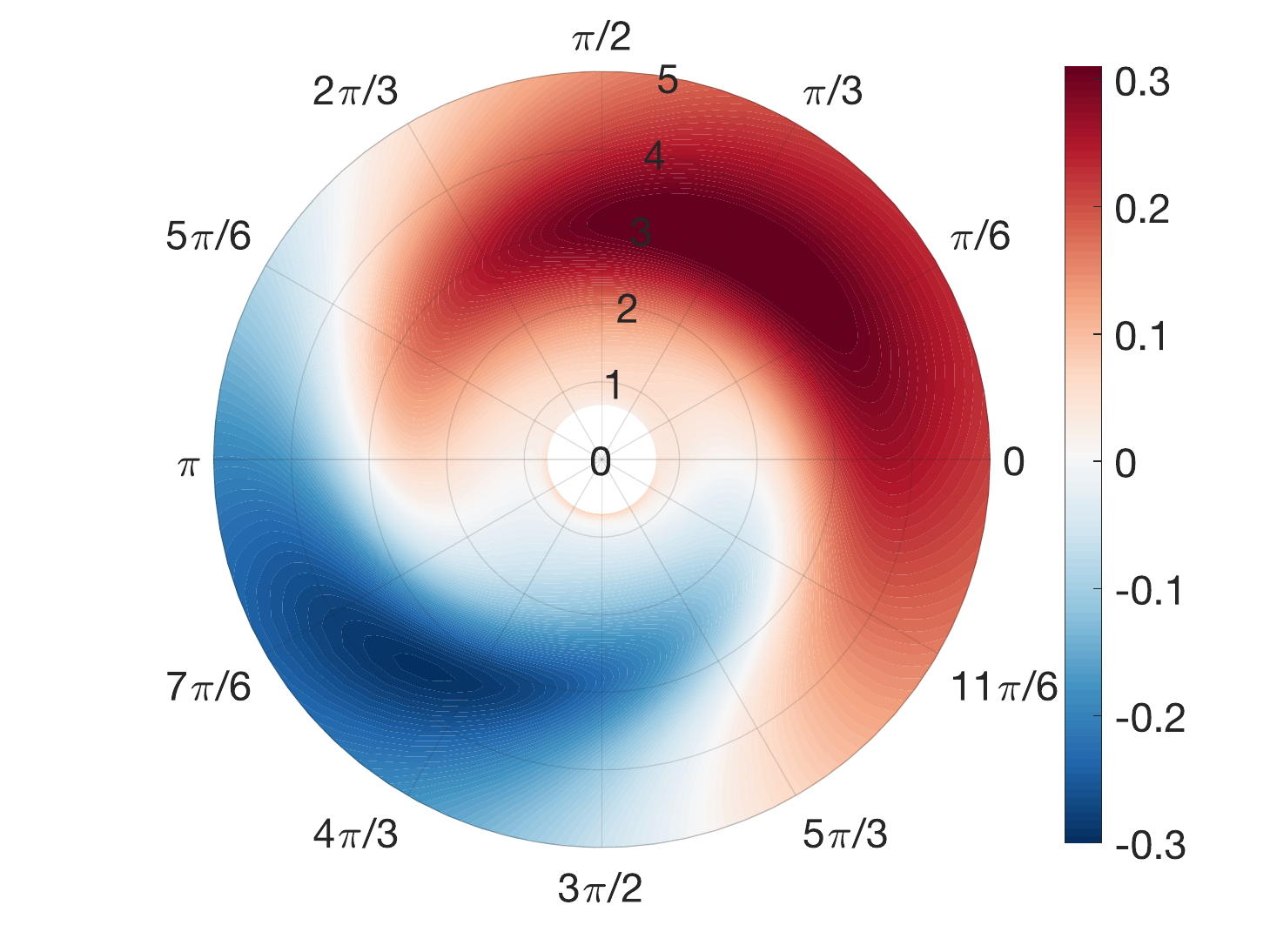}
  \caption{\small $\log{C_T/C_{T_{S_0=0}}}$ \\
  $h = 0, \theta = 0.5$}
\end{subfigure}%
\begin{subfigure}{.5\textwidth}
  \centering
  \includegraphics[width=1\linewidth]{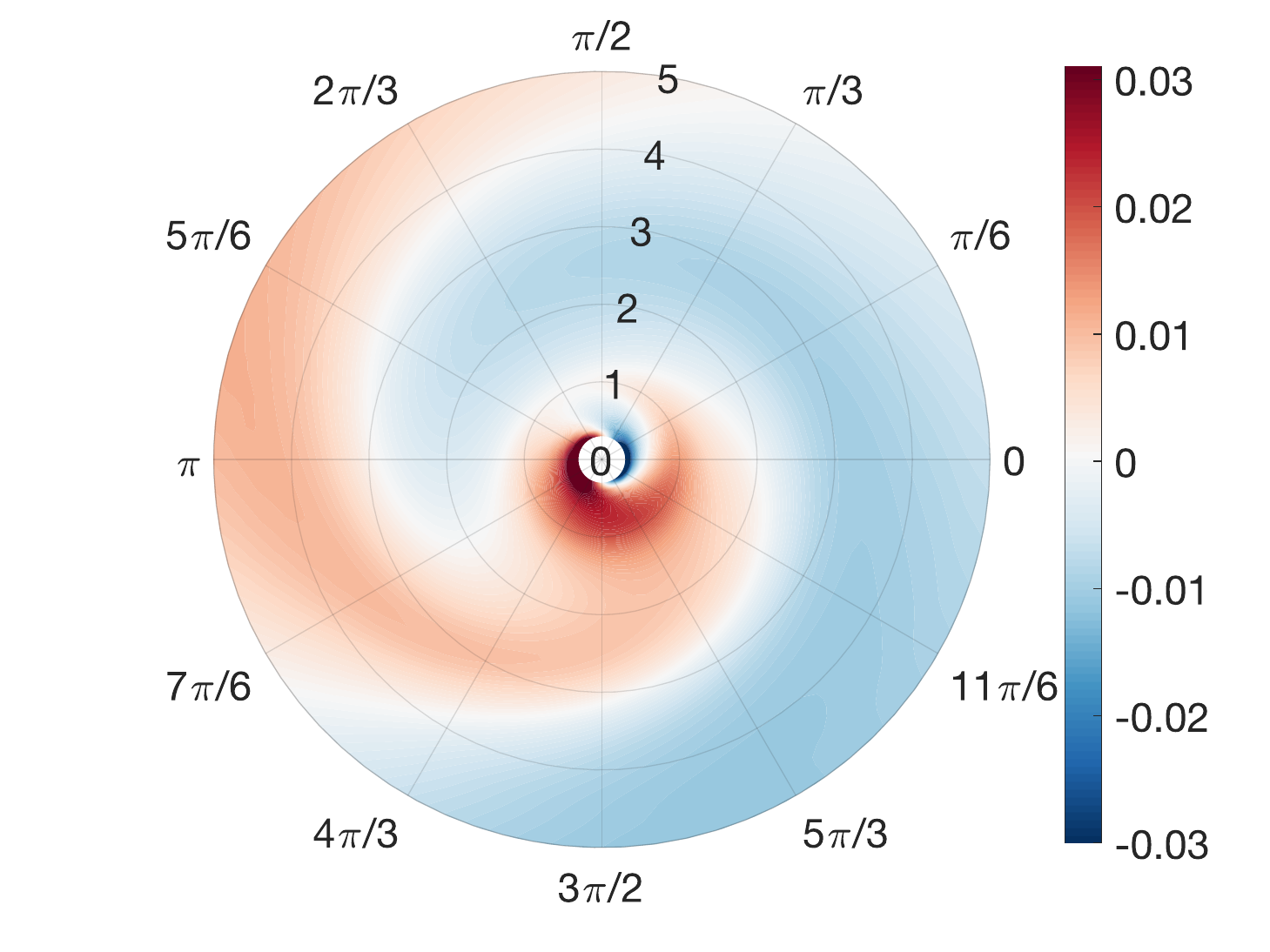}
  \caption{$\Delta \eta$ \\
  $h = 0, \theta = 0.5$}
\end{subfigure} \\
    \begin{subfigure}{.5\textwidth}
  \centering
  \includegraphics[width=1\linewidth]{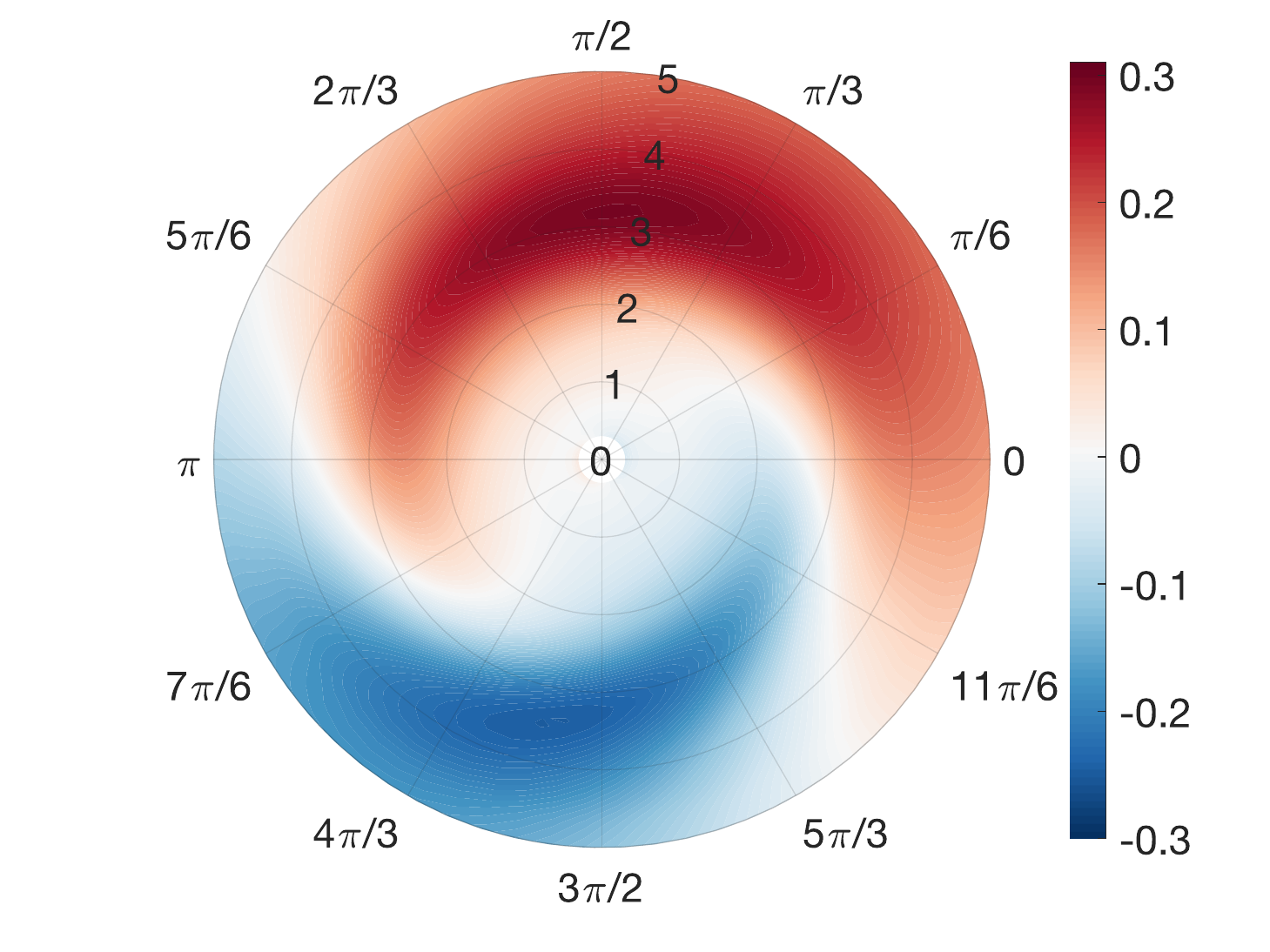}
   \caption{\small $\log{C_T/C_{T_{S_0=0}}}$ \\
  $h = 1, \theta = 0.5$}
\end{subfigure}%
\begin{subfigure}{.5\textwidth}
  \centering
  \includegraphics[width=1\linewidth]{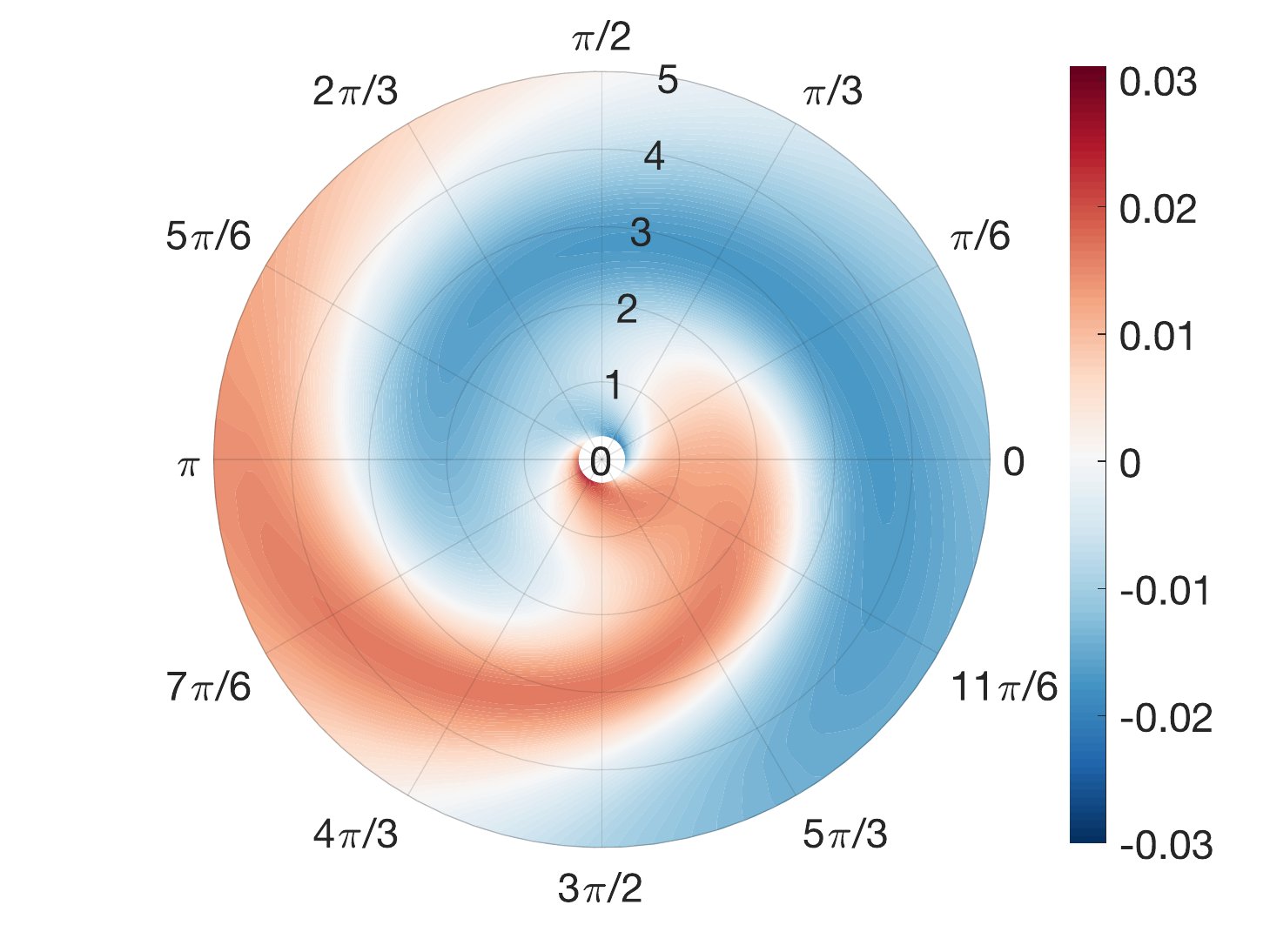}
  \caption{$\Delta \eta$ \\
  $h = 1, \theta = 0.5$}
\end{subfigure}
    \caption{ Impact of stiffness phase offset ($\phi_S$; azimuthal axis) and kinematic frequency ($\sigma$; radial axis) on thrust and efficiency for a pitching (top) and a pitching and heaving (bottom) plate, relative to the constant-stiffness case. Note the white unit circles on the efficiency contours. They are whited out due to large negative efficiency values, which detract from the regions of interest. The parameters used are $\overline S = 20$, $h_0 = 1$, $\theta_0 = -0.5j$, $R = 0.01$, and $|S_0| = 0.5$.}
    \label{fig:polar pitched plots}
\end{figure}

Finally, we consider the role of the phase of stiffness oscillation in more detail. Figure \ref{fig:polar pitched plots} shows the change in thrust and efficiency relative to the constant-stiffness case. 
For the heaving plate, the ideal phase for thrust was approximately between $\pi/2$ and $2\pi/3$; for the pitching plate, however, the ideal phase is approximately $\pi/3$. The ideal phase for the combined pitching and heaving plate is approximately $\pi/2$, directly between the isolated pitching and heaving cases. Generally, the efficiency has opposing behavior to the thrust, with increased thrust coinciding with decreased propulsive efficiency. 


\section{Conclusions}

In this work, we explore the impact of time-periodic stiffness on an oscillating plate in a free stream---a model for swimmers. The stiffness oscillations were varied in amplitude and phase and consistently compared to the baseline constant-stiffness case. The stiffness oscillation frequency is fixed at twice the kinematic frequency to enforce zero mean side force, which is required for rectilinear swimming. The fluid-structure interaction model is solved using a spectral method first introduced in \cite{MOORE2017792}, modified to account for time-periodic stiffness.

It is shown that the oscillation in stiffness has a significant impact on the thrust and kinematics. The impact on thrust is most prominent, with time-periodic stiffness increasing thrust by up to 35\% at the first resonant peak. The impact on efficiency is, on the other hand, mild, with the efficiency changing by $\Delta \eta \le 0.05$. These results align well with \cite{doi:10.1063/5.0027927}. The changes in thrust and efficiency are often negatively correlated. These performance changes are consistent across heaving, pitching, and combined pitch-and-heave motions. 

The performance alteration due to time-varying stiffness is strongly linked to the phase of the stiffness oscillation. This is because thrust and side forces are produced at different stages throughout the kinematic cycle, and whether the plate is more or less stiff at those stages either yields enhanced thrust or power reduction. This is especially evident when comparing pure heaving motions to pure pitching motions, which have different phase differences between the leading edge input and the trailing edge deflection.

When the kinematic frequency and amplitude of stiffness oscillation are large enough, instabilities emerge in regions of parameter space centered about resonant frequencies of the constant-stiffness plate. Our linear method cannot capture the saturation of the instabilities, but we anticipate that the unstable regions of parameter space may yield enhanced propulsive performance. Exploring the unstable regions via experiments or nonlinear simulations is a promising future research direction.

While there may not yet be concrete evidence of biological swimmers actively changing their stiffness on the time scale of their kinematic frequency, we have presented strong evidence that there would be a hydrodynamic benefit to doing so. This may be a promising avenue to pursue when designing robotic swimmers. 

\appendix
\section{Method of solution}
\label{ss: Solution method}
We assume the kinematics $Y$, the hydrodynamic load $q = \Delta p$, and the stiffness ratio $S$ are periodic in time and can be expressed as Fourier series,
\begin{subequations}
    \begin{align}
    Y(x,t)  &= \sum_{m=-\infty}^{\infty} \hat{y}_{m}(x) e^{ j m \sigma t}, \label{eq:fdef}\\
    q(x,t) & = \sum_{m=-\infty}^{\infty}\hat{q}_m(x)e^{ j m \sigma t}, \\
    S(t)  & = \sum_{m=-\infty}^{\infty}\hat{S}_m e^{j m \sigma t}.
    \end{align}
    \label{ss: solution method: Fourier Series}
\end{subequations}
Substituting these expressions into~\eqref{Non-dimensional equation of motion} gives
\begin{equation}
    \sum_{m=-\infty}^{\infty} \hat{q}_m(x) e^{j m \sigma t} + 2R \sum_{m=-\infty}^{\infty}m^2 \sigma ^2 \hat{y}_m (x)e^{j m \sigma t} -  \frac{2}{3}\sum_{m=-\infty}^{\infty}\sum_{k=-\infty}^{\infty} \hat{S}_{m - k} \hat{y}_k''''(x) e^{j m \sigma t} = 0,
\end{equation}
where a prime denotes differentiation with respect to $x$. Separating the Fourier components yields the following system of ordinary differential equations,
\begin{equation}
        \label{eq:ode}
        \hat{q}_m (x) + 2m^2 \sigma^2 R \hat{y}_m (x) -  \frac{2}{3}\sum_{k=-\infty}^{\infty} \hat{S}_{m-k} \hat{y}''''_{k} (x) = 0, \quad \forall m \in \mathbb{Z}.
\end{equation}
Expanding the heaving and pitching motions into Fourier series as
\begin{subequations}
\begin{align}
    h(t) = \sum_{m=-\infty}^{\infty} \hat{h}_m e^{j m \sigma t}, \\
    \theta(t) = \sum_{m=-\infty}^{\infty} \hat{\theta}_m e^{j m \sigma t},
\end{align}
\end{subequations}
and substituting them and~\eqref{eq:fdef} into~\eqref{eq:bcog} gives boundary conditions for the $\hat y_m$:
\begin{equation}
        \label{eq:bcfour}
        \hat{y}_m(-1)  = \hat h_m, \;\;\; \hat{y}'_m(-1) = \hat{\theta}_m, \;\;\; \hat{y}''_m(1) = \hat{y}'''_m(1) = 0, \quad \forall m \in \mathbb{Z}.
\end{equation}
We also require that all the functions are real, leading to the reality conditions,
\begin{equation}
    \hat y_m(x) = \hat y_{-m}^*(x),\; \hat q_m(x) = \hat q_{-m}^*(x),\; \hat S_m = \hat S_{-m}^*,\; \hat h_m = \hat h_{-m}^*,\; \hat \theta_{m} = \hat \theta_{-m}^{*}, \quad \forall m \in \mathbb{Z},
\end{equation}
where the superscript $*$ again denotes complex conjugation. 

Given the hydrodynamic load $q$, the stiffness ratio $S$, and the heaving and pitching actuation at the leading edge, we can solve for the kinematics $Y$ by solving the system of coupled ordinary differential equations given by~\eqref{eq:ode} along with their boundary conditions given by~\eqref{eq:bcfour}. What remains is to couple the solid and fluid mechanics by expressing the hydrodynamic load in terms of the kinematics, which we do next. 

The pressure difference across the plate creates the hydrodynamic load, and is related to Prandtl's acceleration potential by
\begin{equation*}
\Delta p = \phi_\text{top} - \phi_\text{bottom}.
\end{equation*}
Taking the divergence of~\eqref{eq:ndmom} and using incompressibility shows that $\phi$ is harmonic, implying the existence of a harmonic conjugate $\psi$. We define the complex acceleration potential
\begin{equation}
    F(x,y,t) = \phi(x,y,t) + i \psi(x,y,t),
\end{equation}
where $i = \sqrt{-1}$. It will be convenient to work with the complex variable $z = x + iy$. Note that $F$ is analytic in $z$. We conformally map the physical $z$ plane to the exterior of the unit disk in the $\zeta$ plane using
\begin{equation}
    z = \frac{1}{2}\left( \zeta + \frac{1}{\zeta}\right).
    \label{JT}
\end{equation}
Because the acceleration potential $F$ is conformally invariant under $\ref{JT}$, $F$ is analytic in $\zeta$ and can be represented by a multipole expansion,
\begin{equation}
    F = i\left(\frac{a_0(t)}{\zeta + 1} + \sum_{k = 1}^{\infty}\frac{a_k(t)}{\zeta ^k} \right),
    \label{eq:mpole}
\end{equation}
where the coefficients $a_k$ are real \citep{wu_1961}. The first term represents the singularity at the leading edge, and the infinite series represents an analytic function that is regular on and outside the unit circle, decaying in the far field \citep{wu_1961}.

Continuing with our assumption of a time-periodic flow, we expand the coefficients in the multipole expansion into Fourier series,
\begin{equation}
    a_k(t) = \sum_{m=-\infty}^{\infty} \hat{a}_{k,m}e^{j m \sigma t}, \quad \forall k \in \mathbb{W}.
    \label{eqn 2}
\end{equation}
Substituting~\eqref{eqn 2} into~\eqref{eq:mpole} yields
\begin{equation}
    F = i\left(\frac{1}{\zeta + 1}\sum_{m=-\infty}^{\infty} \hat{a}_{0,m}e^{j m \sigma t} + \sum_{k = 1}^{\infty}\frac{1}{\zeta ^k} \sum_{m=-\infty}^{\infty} \hat{a}_{k,m}e^{j m \sigma t}\right),
\end{equation}
which we can rewrite as a Fourier expansion of $F$,
\begin{equation}
    \label{eq:fpot}
    F = \sum_{m = -\infty}^{\infty} \hat F_m e^{jm\sigma t} = \sum_{m=-\infty}^{\infty} i\left(\frac{\hat{a}_{0,m}}{\zeta + 1} + \sum_{k = 1}^{\infty}\frac{\hat{a}_{k,m}}{\zeta ^k} \right) e^{jm\sigma t}.
\end{equation}
Evaluating~\eqref{eq:fpot} on the unit circle ($\zeta = e^{i \theta}$), which corresponds to the surface of the plate in the $z$ plane, and separating the real and imaginary parts yields
\begin{subequations}
\begin{align}
    \phi|_W &= \sum_{m=-\infty}^{\infty} \left(\frac{1}{2}\hat{a}_{0,m} \tan \frac{\theta}{2} + \sum_{k = 1}^\infty \hat{a}_{k,m} \sin{k\theta} \right) e^{j m \sigma t}, \\
    \psi|_W &= \sum_{m=-\infty}^{\infty} \left( \frac{1}{2} \hat a_{0,m} + \sum_{k = 1}^{\infty} \hat{a}_{k,m}\cos{k\theta}\right) e^{j m \sigma t} = \sum_{m=-\infty}^{\infty} \Psi_m(x) e^{j m \sigma t}.
    \label{ss: solution method: complex conjugate of pressure}
\end{align}
\end{subequations}
We recognize the expression for $\Psi_m(x)$ as a Chebyshev series in $x$,
\begin{equation}
    \Psi_{m}(x) = \frac{1}{2}\hat{a}_{0,m} + \sum_{k = 1}^{\infty} \hat{a}_{k,m} T_{k} (x), \quad \forall m \in \mathbb{Z},
\end{equation}
where $T_k(x) = \cos(k \arccos x)$ is the $k$th Chebyshev polynomial, and $x = \cos \theta$.

The complex acceleration potential can be related to the complex velocity $w = u - iv$ through the momentum equation,~\eqref{eq:ndmom}, by
\begin{equation}
    \frac{\partial F}{\partial z} = \frac{\partial w}{\partial t} +  \frac{\partial w}{\partial z}.
\end{equation}
Evaluating the imaginary part on the plate's surface ($z = x$) and substituting the no-penetration boundary condition,~\eqref{eq:npbc}, yields
\begin{equation}
    \frac{\partial \psi}{\partial x}\Bigr|_{y=0} = -\left(\frac{\partial}{\partial t} + \frac{\partial}{\partial x} \right) ^{2} Y.
    \label{ss:solution method: No pen BC}
\end{equation}
Substituting~\eqref{eq:fdef} and~\eqref{ss: solution method: complex conjugate of pressure} into~\eqref{ss:solution method: No pen BC} gives
\begin{equation}
    \sum_{m=-\infty}^{\infty} \mathcal{D} \Psi_m (x) e^{j m \sigma t} = -\sum_{m=-\infty}^{\infty} (j m \sigma + \mathcal{D})^{2}\hat{y}_m (x)e^{j m \sigma t},
\end{equation}
where $\mathcal D = \frac{\text{d}}{\text{d}x}$. Separating Fourier components gives
\begin{equation}
    \label{eq:npbcf}
    \mathcal{D}\Psi_m (x) = -(j m \sigma + \mathcal{D})^{2} \hat{y}_m (x), \quad \forall m \in \mathbb{Z}.
\end{equation}
Given $\hat y_m$, we expand it into a Chebyshev series and use~\eqref{eq:npbcf} to express the Chebyshev coefficients of $\Psi_m$---that is, $\hat a_{k,m}$ for $k \ge 1$---in terms of the Chebyshev coefficients of $\hat y_m$. 
To determine $\hat{a}_{0,m}$, we expand the vertical velocity on the plate in a Fourier series,
\begin{equation}
    v|_{W} = \sum_{m = -\infty}^{\infty} \hat{v}_m(x) e^{j m \sigma t},
\end{equation}
and the spatial coefficients in Chebyshev series,
\begin{equation}
    \hat{v}_m(x) = \frac{1}{2} \hat{V}_{0,m} + \sum_{k = 1}^{\infty} \hat{V}_{k,m}T_k(x), \quad \forall m \in \mathbb{Z}.
\end{equation}
We can express the no-penetration boundary condition,~\eqref{eq:npbc}, as
\begin{equation}
    \label{eq:vvel}
    \hat{v}_m(x) = (j m \sigma + \mathcal{D})\hat{y}_m(x), \quad \forall m \in \mathbb{Z}.
\end{equation}
Given the Chebyshev coefficients of $\hat y_m$,~\eqref{eq:vvel} gives the Chebyshev coefficients of $\hat v_m$. The coefficient $\hat a_{0,m}$ is then given by
\begin{equation}
    \hat{a}_{0,m} = -C(j m\sigma)(\hat{V}_{0,m} + \hat{V}_{1,m}) + \hat{V}_{1,m}, \quad \forall m \in \mathbb{Z},
\end{equation}
where 
\begin{equation}
    C(j m \sigma) = \frac{K_1(jm\sigma)}{K_0(jm\sigma) + K_1(jm\sigma)}
\end{equation}
is the Theodorsen function, and $K_\nu$ is the modified Bessel function of the second kind of order $\nu$. The expression for $\hat a_{0,m}$ is derived in \citet{wu_1961}. 

With all the $\hat a_{k,m}$ in hand, the hydrodynamic load is given by
\begin{equation}
    \hat q_m(x) = \hat a_{0,m} \sqrt{\frac{1-x}{1+x}} + 2 \sum_{k=1}^{\infty} \hat a_{k,m} \sin k\theta, \quad \forall m \in \mathbb{Z}.
\end{equation}
The hydrodynamic load $\hat q_m$ depends linearly on $\hat y_m$. 

To summarize, given the kinematics $\hat y_m$, we can calculate the coefficients $\hat a_{k,m}$, with which we can calculate the hydrodynamic load, which alters the kinematics via~\eqref{eq:ode}. The coupled fluid-structure problem must be solved numerically. The pseudospectral numerical method for constant stiffness is given by \citet{MOORE2017792}, which is relatively straightforward to adapt to account for time-periodic stiffness; in it, the kinematics are expanded into Chebyshev series. All infinite series are truncated to finite series; for the results presented in this work, we have used 16 Fourier modes and 64 Chebyshev modes. The method is fast and accurate, pre-conditioning the system with continuous operators to avoid errors typically encountered when using spectral methods to solve high-order differential equations. Formulas to calculate the thrust and power coefficients are given in \citet{MOORE2017792}.

\section{Floquet analysis}
\label{s:floquet theory}

To test the stability of the solutions computed using the method in Appendix~\ref{ss: Solution method}, we perform a Floquet analysis of the problem with homogeneous boundary conditions. This analysis is adapted from the eigenvalue problem described by \citet{floryan_rowley_2018, floryan_rowley_2020}. Following the preceding analysis, but not assuming a form for the time dependence, we arrive at 
\begin{subequations}
    \begin{gather}
        2R Y_{tt} + \frac{2}{3} S(t) Y_{xxxx} = \Delta p, \\
        Y(x,t) = \frac{1}{2} y_0(t) + \sum_{k=1}^\infty y_k(t) T_k(x), \label{eq:ycheb}\\
        \Delta p(x,t) = a_0(t)\sqrt{\frac{1-x}{1+x}}+2\sum_{k=1}^{\infty}a_k(t)\sin{k \theta}, \\
        \sum_{k=1}^{\infty}a_k(t) T_k'(x) = -\frac{1}{2}\ddot{y}_0(t)  - \sum_{k=1}^{\infty}[ \ddot{y}_k(t) T_k (x) + 2 \dot{y}_k(x) T_k'(x) + y_k (t)T_k''(x)], \label{eq:flonp} \\
    Y(-1,t) = 0, \quad Y_{x}(-1,t)  = 0, \quad Y_{xx}(1,t) = Y_{xxx}(1,t) = 0. \label{eq:bcflo}
    \end{gather}
\end{subequations}
Above, a dot denotes differentiation with respect to $t$. In~\eqref{eq:ycheb}, we have written $Y$ as a Chebyshev series in $x$. The expression in~\eqref{eq:flonp} derives from the no-penetration condition~\eqref{ss:solution method: No pen BC}. 

In what follows, we set $S(t) =  \overline S + \frac{\overline S S_0}{2}( e^{ j \omega t} + e^{- j \omega t})$, as in~\eqref{eq:stif}. Here, $\omega$ is the frequency of the stiffness oscillation. Assuming the Floquet solution form,
\begin{subequations}
    \begin{align}
    y_k(t) &= e^{\lambda t} \sum_{m = -\infty}^{\infty} \hat y_{k,m}e^{ j m \omega t}, \label{eq:yflo} \\
    a_k(t) &= e^{\lambda t} \sum_{m = -\infty}^{\infty} \hat a_{k,m} e^{ j m \omega t}, \quad \forall k \in \mathbb{W},
    \end{align}
\end{subequations}
where $\lambda$ is the Floquet exponent, gives, after separating the Fourier components, the following set of equations,
\begin{subequations}
    \begin{gather}
        2R(\lambda + j m \omega)^2 \bbeta_m + \frac{2}{3} \overline{S} \Dfour \bbeta_m  + \frac{1}{3} \overline S S_0  \Dfour \bbeta_{m-1} + \frac{1}{3} \overline S S_0 \Dfour \bbeta_{m+1} = \hat{\boldsymbol p}_m, \\
        \hat{\boldsymbol p}_m = \mathbf A \hat{\boldsymbol a}_m, \label{eq:palin}\\
        \mathbf D \hat{\boldsymbol a}_m = -(\lambda + j m \omega)^2 \bbeta_m - 2(\lambda + j m \omega) \mathbf{D} \bbeta_m - \mathbf{D}^2 \bbeta_m, \\ 
        \hat{\boldsymbol V}_m = (\lambda + j m \omega) \bbeta_m + \mathbf{D}\bbeta_m, \\
        \hat a_{0,m} = -C(\lambda + j m \omega)(\hat V_{0,m} + \hat V_{1,m}) + \hat V_{1,m}, \quad \forall m \in \mathbb{Z}.
    \end{gather}
    \label{First system of equations for Floquet analysis}
\end{subequations}
Above, $\bbeta_m$ is a vector of the Chebyshev coefficients corresponding to index $m$ in~\eqref{eq:yflo}: $\bbeta_m = \begin{bmatrix} \hat y_{0,m} & \hat y_{1,m} & \hat y_{2,m} & \cdots \end{bmatrix}^T$; analogous expressions hold for $\hat{\boldsymbol p}_m$ (pressure), $\hat{\boldsymbol a}_m$ (potential), and $\hat{\boldsymbol V}_m$ (vertical velocity).  The equality in~\eqref{eq:palin} states that the Chebyshev coefficients of the pressure are linear combinations of the Chebyshev coefficients of the potential, and $\mathbf{D}$ is the differentiation operator in Chebyshev space. Putting these equations together gives
\begin{equation}
\begin{split}
    \mathbf{A}[\mathbf{e_1}(\mathbf{e_2}-C(\lambda_m)\mathbf{e_1}-C(\lambda_m)\mathbf{e_2})^{T}(\lambda_m \boldsymbol{\hat y}_m + \mathbf{D} \boldsymbol{\hat y}_m) - \lambda_m ^2 \mathbf{D^{-}} \boldsymbol{\hat y}_m  - 2 \lambda_m \mathbf{D^{-}} \mathbf{D}\boldsymbol{\hat y}_m \\ - \mathbf{D^{-}}\mathbf{D^2} \boldsymbol{\hat y}_m] - 2R\lambda_m^2 \boldsymbol{\hat y}_m - \frac{2}{3}\mathbf{D^{4}} \overline{S} \boldsymbol{\hat y}_m = \frac{\overline S S_0}{3}\mathbf{D}^4 [ \boldsymbol{\hat y}_{m-1} + \boldsymbol{\hat y}_{m+1}], \quad \forall m \in \mathbb{Z},
    \end{split}
    \label{First eom for Floquet analysis}
\end{equation}
where $\lambda_m = \lambda + j m \omega$, $\mathbf{D^{-}}$ is the Chebyshev-space representation of the integration operator that makes the first Chebyshev coefficient zero, and $\mathbf{e}_k$ is the $k$th Euclidean basis vector. 

Because of the presence of the Theodorsen function, solving for the Floquet exponents requires solving a nonlinear generalized eigenvalue problem. Since we are mainly interested in delineating regions of parameter space where solutions are unstable, we instead follow the idea of \citet{kumar_tuckerman_1994} to find the marginal stability curves. Rather than calculating the Floquet exponents for given values of the parameters, we set the value of the Floquet exponent such that $Re(\lambda) = 0$ and solve for $S_0$. Physically, we are trying to find the strength of the stiffness oscillation that borders regions of stability and instability. This leads to a linear generalized eigenvalue problem where $S_0$ is the eigenvalue. Note that we must also set a value for $Im(\lambda) \in [0, \omega]$; $Im(\lambda) = 0$ is called the harmonic case, and $Im(\lambda) = \omega/2$ is called the subharmonic case \citep{kumar_tuckerman_1994}.

As can be seen in~\eqref{First eom for Floquet analysis}, the time dependence of the stiffness causes cross-frequency coupling in the kinematics. We write~\eqref{First eom for Floquet analysis} compactly as
\begin{equation}
    \mathbf{B}_m \boldsymbol{\hat y}_{m} = \frac{\overline S S_0}{3} \mathbf{C}_m\boldsymbol{\hat y}_{m}, \quad \forall m \in \mathbb{Z},
    \label{Inital linear system}
\end{equation}
where,
\begin{align*}
    \mathbf{B}_m =&    \mathbf{A}[\mathbf{e_1}(\mathbf{e_2}-C(\lambda_m)\mathbf{e_1}-C(\lambda_m)\mathbf{e_2})^{T}(\lambda_m \mathbf{I} + \mathbf{D})  - \lambda_m ^2 \mathbf{D^{-}}   - 2 \lambda_m \mathbf{D^{-}} \mathbf{D} - \mathbf{D^{-}}\mathbf{D^2}] \\
    &- 2R \lambda_m^2 \mathbf{I} - \frac{2}{3}\boldsymbol{D^{4}} \overline{S}, \quad \forall m \in \mathbb{Z}.
\end{align*}
The operator $\mathbf C_m $ maps $\boldsymbol{\hat y}_m$ to $\mathbf{D}^4 (\boldsymbol{\hat y}_{m+1} +\boldsymbol{\hat y}_{m-1})$. This is a linear generalized eigenvalue problem where the eigenvalues are the amplitudes of the stiffness oscillation, $S_0$. 

To proceed, we truncate all Chebyshev expansions to the upper limit $N$, and truncate all Fourier expansions so that they contain frequencies up to $M\omega$. Doing so makes $\boldsymbol{\hat y}_m$ a vector of length $N+1$. We must simultaneously solve the system~\eqref{Inital linear system} for all $\boldsymbol{\hat y}_m$, leading to a system of size $2M(N+1) \times 2M(N+1)$. We choose $N = 16$ and $M = 11$. We reduce the order of the Chebyshev and Fourier modes so the Floquet problem can be solved in reasonable time. 

In the harmonic ($Im(\lambda) = 0$) and subharmonic ($Im(\lambda) = \omega$) cases, $\boldsymbol{\hat y}_m$ must obey the reality conditions: $\boldsymbol{\hat y}_{-m} = \boldsymbol{\hat y}_m^*$ in the harmonic case, and $\boldsymbol{\hat y}_{-m} = \boldsymbol{\hat y}_{m-1}^*$ in the subharmonic case. The reality conditions allow us to rewrite the Fourier expansions in terms of only non-negative indices. For $0 < Im(\lambda) < \omega$, positive and negative frequencies are independent of each other, and~\eqref{eq:yflo} must be added to its complex conjugate to form a real field. 

Explicitly writing the real and imaginary components in~\eqref{Inital linear system}  yields
\begin{equation*}
    (\mathbf{B}_m^r + i\mathbf{B}_m^i) (\boldsymbol{\hat y}_m^r + i\boldsymbol{\hat y}_m^i) = \frac{\overline S S_0}{3} \mathbf{D}^4 ( \boldsymbol{\hat y}_{m-1}^r + i\boldsymbol{\hat y}_{m-1}^i + \boldsymbol{\hat y}_{m+1}^r + i\boldsymbol{\hat y}_{m+1}^i), \quad \forall m \in \mathbb{Z},
\end{equation*}
from which we can separate the real and imaginary components to get
\begin{subequations}
    \begin{gather}
        \mathbf{B}_m^r \boldsymbol{\hat y}_m^r - \mathbf{B}_m^i \boldsymbol{\hat y}_m^i = \frac{\overline S S_0}{3} \mathbf{D}^4 (\boldsymbol{\hat y}_{m-1}^r + \boldsymbol{\hat y}_{m+1}^r), \\
        \mathbf{B}_m^r \boldsymbol{\hat y}_m^i + \mathbf{B}_m^i \boldsymbol{\hat y}_m^r = \frac{\overline S S_0}{3} \mathbf{D}^4 (\boldsymbol{\hat y}_{m-1}^i + \boldsymbol{\hat y}_{m+1}^i), \quad \forall m \in \mathbb{Z}. 
    \end{gather}
    \label{real and imaginary parts split}
\end{subequations}
This can be rewritten in matrix form as 
\begin{equation}
    \begin{bmatrix}
      \mathbf{B}_m^r & - \mathbf{B}_m^i \\
      \mathbf{B}_m^i & \mathbf{B}_m^r
    \end{bmatrix}
    \begin{bmatrix}
      \boldsymbol{\hat y}_m^r \\
      \boldsymbol{\hat y}_m^i 
    \end{bmatrix} = \frac{\overline S S_0}{3} \begin{bmatrix}
      \mathbf{C}_m & \mathbf{0} \\
      \mathbf{0} & \mathbf{C}_m
    \end{bmatrix}
    \begin{bmatrix}
        \boldsymbol{\hat y}_m^r \\
        \boldsymbol{\hat y}_m^i 
    \end{bmatrix}, \quad \forall m \in \mathbb{Z},
    \label{matrix form of real and imag parts split}
\end{equation}
where $\mathbf C_m $ is as before. 

Finally, we incorporate the boundary conditions into~\eqref{matrix form of real and imag parts split}. The formula to evaluate a Chebyshev series at the endpoints is
\begin{equation*}
    \hat y_m(\pm 1) = \frac{1}{2} {\hat y}_{0,m} + \sum_{k = 1} ^N (\pm 1)^k {\hat y}_{k,m},
\end{equation*}
which can be split into its real and imaginary parts,
\begin{subequations}
    \begin{gather}
         \hat y_m^r(\pm 1) = \frac{1}{2} {\hat y}_{0,m}^r  +\sum_{k = 1} ^N (\pm 1)^k {\hat y}_{k,m}^r, \\
          \hat y_m^i(\pm 1) = \frac{1}{2} {\hat y}_{0,m}^i  +\sum_{k = 1} ^N (\pm 1)^k {\hat y}_{k,m}^i.
    \end{gather}
    \label{BCs}
\end{subequations}
This allows us to express the boundary conditions in~\eqref{eq:bcflo} in terms of the Chebyshev coefficients. To enforce the boundary conditions, we replace the last four rows of the first block in~\eqref{matrix form of real and imag parts split} by the four boundary conditions on the real part, and the last four rows in~\eqref{matrix form of real and imag parts split} by the four boundary conditions on the imaginary part. Combining the equations for all values of $m$ into one large system leads to a generalized eigenvalue problem of the form $(3/\overline S)\mathbf B \boldsymbol{\hat y} = S_0 \mathbf{C_H} \boldsymbol{\hat y}$, which is linear in $S_0$. In the harmonic case, the matrices $\mathbf B$ and $\mathbf C_{H}$ take the form
\begin{subequations}
    \begin{gather}
        \mathbf{B} = \begin{bmatrix}
          \mathbf{B}_0^r & - \mathbf{B}_0^i & \boldsymbol 0 & \boldsymbol 0 & \boldsymbol 0 & \boldsymbol 0 & \cdots \\  
          \mathbf{B}_0^i & \mathbf{B}_0^r  & \boldsymbol 0 & \boldsymbol 0 & \boldsymbol 0 & \boldsymbol 0 & \cdots \\ 
             \boldsymbol 0 & \boldsymbol 0 &\mathbf{B}_1^r & - \mathbf{B}_1^i & \boldsymbol 0 & \boldsymbol 0 & \cdots   \\
             \boldsymbol 0 & \boldsymbol 0 & \mathbf{B}_1^i & \mathbf{B}_1^r & \boldsymbol 0 & \boldsymbol 0 & \cdots   \\ 
            \boldsymbol 0 & \boldsymbol 0 & \boldsymbol 0 & \boldsymbol 0  &\mathbf{B}_2^r & - \mathbf{B}_2^i & \cdots \\
            \boldsymbol 0 & \boldsymbol 0 & \boldsymbol 0 & \boldsymbol 0  & \mathbf{B}_2^i & \mathbf{B}_2^r & \cdots \\
            \mathbf{0} & \mathbf{0} &\mathbf{0} & \mathbf{0} &\mathbf{0} & \mathbf{0} & \hdots \\
            \mathbf{0} & \mathbf{0} &\mathbf{0} & \mathbf{0} &\mathbf{0} & \mathbf{0} & \hdots \\
            \vdots & \vdots & \vdots & \vdots  & \vdots & \vdots & \ddots
        \end{bmatrix}, \\
    \mathbf{C_H} = 
   \begin{bmatrix}
    \mathbf{0}& \mathbf{0} & 2 \mathbf{D}^4  & \mathbf{0} & \mathbf{0} & \mathbf{0} & \cdots \\
   \mathbf{0} & \mathbf{0} & \mathbf{0} & \mathbf{0} & \mathbf{0} & \mathbf{0} & \cdots \\
   \mathbf{D}^4  & \mathbf{0} & \mathbf{0} & \mathbf{0} & \mathbf{D}^4 & \mathbf{0} & \cdots \\
   \mathbf{0} & \mathbf{D}^4 & \mathbf{0} & \mathbf{0} & \mathbf{0} & \mathbf{D}^4 & \cdots \\
   \mathbf{0} & \mathbf{0} & \mathbf{D}^4 & \mathbf{0} & \mathbf{0} & \mathbf{0} & \cdots \\
   \mathbf{0} & \mathbf{0} & \mathbf{0} & \mathbf{D}^4 & \mathbf{0} & \mathbf{0} & \cdots \\
   \mathbf{0} & \mathbf{0} & \mathbf{0} & \mathbf{0} & \mathbf{D}^4 & \mathbf{0} & \cdots \\ 
   \vdots & \vdots & \vdots & \vdots & \vdots & \vdots & \ddots
   \end{bmatrix},
    \end{gather}
\end{subequations}
with the boundary conditions incorporated as previously described, and the Chebyshev coefficients for all frequencies are stored in the vector
\begin{equation*}
    \boldsymbol{\hat y} = \begin{bmatrix}  
    (\boldsymbol{\hat y}_0^r)^T &
    (\boldsymbol{\hat y}_0^i)^T &
    (\boldsymbol{\hat y}_1^r)^T &
    (\boldsymbol{\hat y}_1^i)^T & 
    \hdots &
    (\boldsymbol{\hat y}_M^r)^T &
    (\boldsymbol{\hat y}_M^i)^T
    \end{bmatrix}^T.
\end{equation*}

For the subharmonic case, $\mathbf B$ is identical to the harmonic case, but by using the reality condition $\boldsymbol{\hat y}_{-m} = \boldsymbol{\hat y}_{m-1}^*$,  $\mathbf C_{SH}$ takes the form
\begin{equation}
        \mathbf{C}_{SH} = 
   \begin{bmatrix}
    \mathbf{D}^4 & \mathbf{0} &  \mathbf{D}^4  & \mathbf{0} & \mathbf{0} & \mathbf{0} & \cdots \\
   \mathbf{0} & -\mathbf{D}^4 & \mathbf{0} & \mathbf{D}^4 & \mathbf{0} & \mathbf{0} & \cdots \\
   \mathbf{D}^4  & \mathbf{0} & \mathbf{0} & \mathbf{0} & \mathbf{D}^4 & \mathbf{0} & \cdots \\
   \mathbf{0} & \mathbf{D}^4 & \mathbf{0} & \mathbf{0} & \mathbf{0} & \mathbf{D}^4 & \cdots \\
   \mathbf{0} & \mathbf{0} & \mathbf{D}^4 & \mathbf{0} & \mathbf{0} & \mathbf{0} & \cdots \\
   \mathbf{0} & \mathbf{0} & \mathbf{0} & \mathbf{D}^4 & \mathbf{0} & \mathbf{0} & \cdots \\
   \mathbf{0} & \mathbf{0} & \mathbf{0} & \mathbf{0} & \mathbf{D}^4 & \mathbf{0} & \cdots \\ 
   \vdots & \vdots & \vdots & \vdots & \vdots & \vdots & \ddots
   \end{bmatrix}.
\end{equation}

{\section{Code validation}

We modify the fast Chebyshev method introduced by 
\cite{MOORE2017792} to account for time-periodic stiffness. We validated the results produced from our code with the plots found in the previous papers. Due to the Fourier series expansion method and the time-periodic stiffness used here we have cross coupling of Fourier modes. When the plate has constant stiffness in time there is no coupling of modes; in fact, only the frequency that the plate is actuated at is active. We verify that the results produced from our code with constant stiffness is identical to those produced by \cite{MOORE2017792}, up to a tolerance. These plots are seen in figure \ref{Moore_validation}.

\begin{figure}
    \centering
    \begin{subfigure}{.5\textwidth}
  \centering
  \includegraphics[width=1\linewidth]{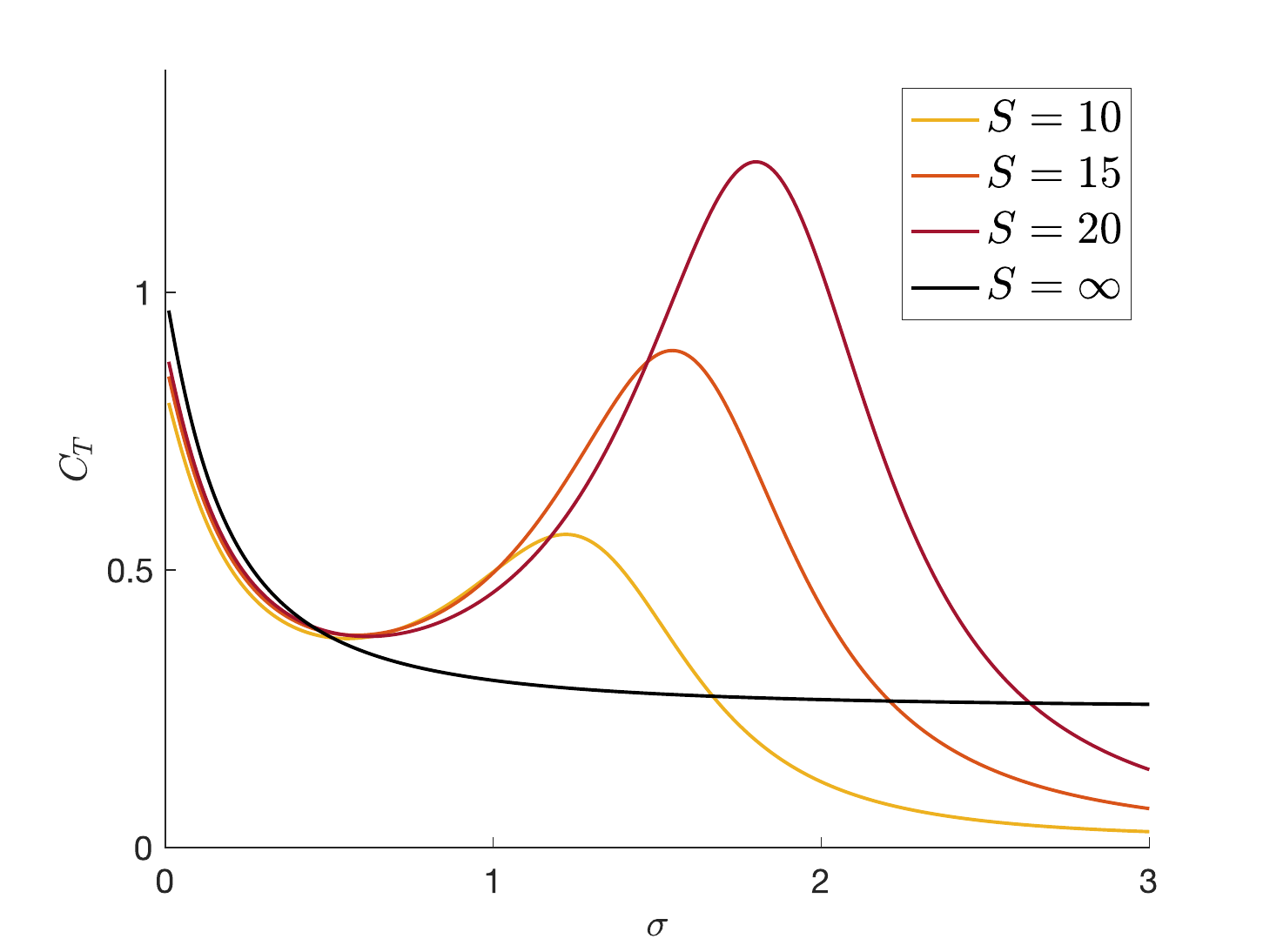}
  \caption{$C_T$}
\end{subfigure}%
\begin{subfigure}{.5\textwidth}
  \centering
  \includegraphics[width=1\linewidth]{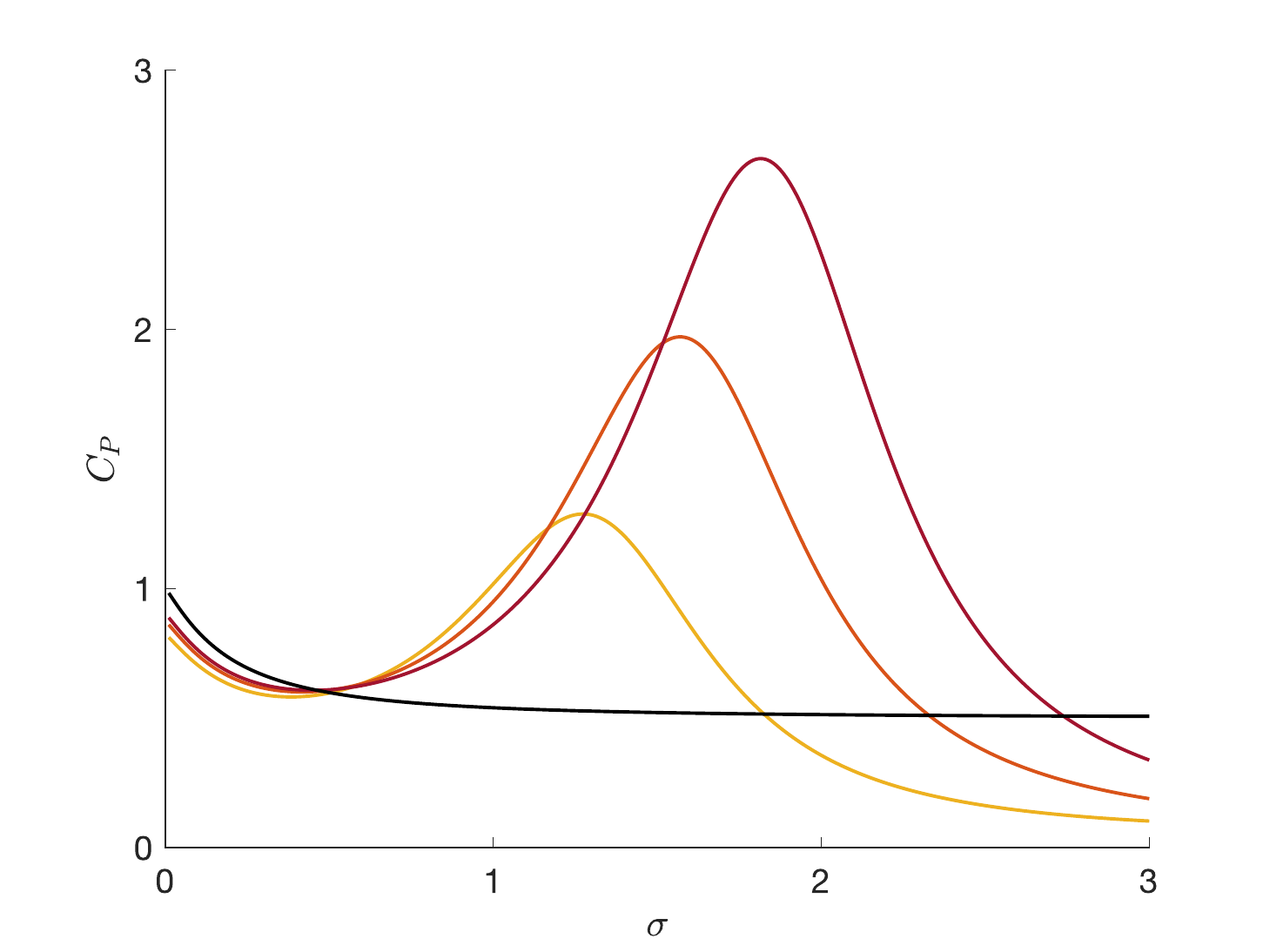}
  \caption{$C_P$}
\end{subfigure} %
\begin{subfigure}{.5\textwidth}
  \centering
  \includegraphics[width=1\linewidth]{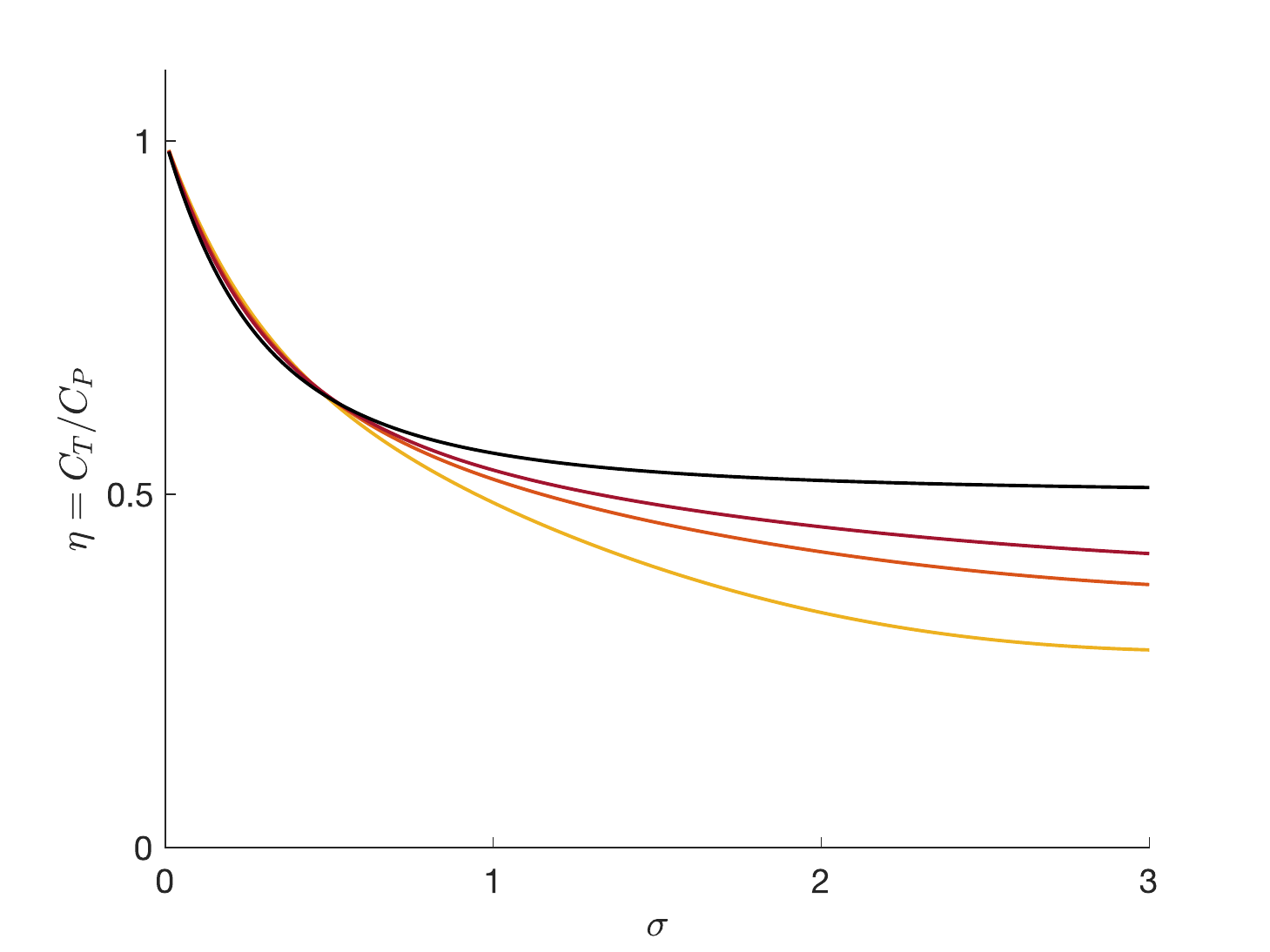}
  \caption{$\eta = C_P/C_T$}
\end{subfigure}%
\caption{{Reproduction of figure 7 in \cite{MOORE2017792}}}
\label{Moore_validation}
\end{figure}
}

\vskip 2mm
\noindent \textbf{Declaration of Interests:} The authors report no conflict of interest.


\newpage
\bibliographystyle{jfm}
\bibliography{jfm}

\end{document}